\documentclass[pdflatex,sn-mathphys-ay]{sn-jnl}

\usepackage{graphicx}%
\usepackage{multirow}%
\usepackage{amsmath,amssymb,amsfonts}%
\usepackage{amsthm}%
\usepackage{mathrsfs}%
\usepackage[title]{appendix}%
\usepackage{xcolor}%
\usepackage{textcomp}%
\usepackage{manyfoot}%
\usepackage{booktabs}%
\usepackage{listings}%
\usepackage{subcaption}
\usepackage{enumitem}
\usepackage{array}
\usepackage{comment}

\begin{document}

\title[Article Title]{A Novel Optical Design for Wide-Field Imaging in X-ray
Astronomy}

\author*[1,2]{\fnm{Neeraj K.} \sur{Tiwari}}\email{neeraj@prl.res.in}

\author[1]{\fnm{Santosh V.} \sur{Vadawale}}\email{santoshv@prl.res.in}

\author[1]{\fnm{N. P. S.} \sur{Mithun}}\email{mithun@prl.res.in}

\affil*[1]{\orgdiv{Astronomy and Astrophysics Division}, \orgname{Physical Research Laboratory}, \orgaddress{\city{Ahmedabad}, \postcode{380009}, \state{Gujarat}, \country{India}}}

\affil[2]{\orgname{Indian Institute of Technology}, \orgaddress{\city{Gandhinagar}, \postcode{382335}, \state{Gujarat}, \country{India}}}

\abstract{Over the decades, astronomical X-ray telescopes have utilized the Wolter type-1 optical design, which provides stigmatic imaging in axial direction but suffers from coma and higher-order aberrations for off-axis sources. The Wolter–Schwarzschild design, with stigmatic imaging in the axial direction, while suffering from higher-order aberrations, is corrected for coma, thus performing better than the Wolter type-1. The Wolter type-1 and Wolter–Schwarzschild designs are optimized for on-axis but have reduced angular resolution when averaged over a wide field of view, with the averaging weighted by the area covered in the field of view. An optical design that maximizes angular resolution at the edge of the field of view rather than at the center is more suitable for wide-field X-ray telescopes required for deep-sky astronomical surveys or solar observations. A Hyperboloid-Hyperboloid optical design can compromise axial resolution to enhance field angle resolution, hence providing improved area-weighted average angular resolution over the Wolter–Schwarzschild design, but only for fields of view exceeding a specific size. Here, we introduce a new optical design that is free from coma aberration and capable of maximizing angular resolution at any desired field angle. This design consistently outperforms Wolter-1, Wolter–Schwarzschild, and Hyperboloid–Hyperboloid designs when averaged over any field of view size. The improvement in performance remains consistent across variations in other telescope parameters such as diameter, focal length, and mirror lengths. By utilizing this new optical design, we also present a design for a full-disk imaging solar X-ray telescope.}


\maketitle

\section{Introduction}
High-sensitivity X-ray observations of astrophysical sources primarily relies on X-ray telescopes developed based on grazing incident optics. Since efficient reflections of X-rays occur only at very grazing incident angles (rays are nearly parallel to the reflecting mirror, with an angle of less than a degree), X-ray telescope designs differ significantly from visible telescopes, where reflection occurs at normal incident angles. In order to increase the reflectivity of X-rays, in addition to the grazing incident angle requirement to achieve total external reflection, the X-ray reflecting mirror surface needs to be made of high Z elements and needs to be polished up to very low surface roughness with an RMS value comparable to X-ray wavelength to reduce scattering \citep{aschenbach1985x}. For hard X-rays (energy $>$ 10 keV), X-ray reflectivity is poor, and it rapidly degrades with further increases in X-ray energy. This can be addressed by obtaining multiple reflections with constructive interference
by having optimally spaced multiple high Z reflecting surfaces. The space between high Z reflecting surfaces is filled with low Z material \citep{mondal2021darpanx, windt2015advancements}.

X-ray telescopes suffer from poor reflectivity, necessitating a minimal reflection optical design. Single-reflection designs are sensitive to coma aberration \citep{pivovaroff2023geometries, giacconi1960telescope}, which can only be controlled with an even number of reflections \citep{burrows1992optimal}. Thus, dual-reflection designs are preferred, as they experience less effective area degradation and design complexity compared to designs with more reflections. The Wolter type-1 ($W1$) is a dual-reflection optical design with axially symmetric confocal mirrors: a paraboloid as the primary and a hyperboloid as the secondary \citep{wolter1952spiegelsysteme, vanspeybroeck1972design}. Both reflections converge toward the optical axis, resulting in a reduced focal length. 

The geometrical effective area of dual-reflection, grazing-incidence X-ray telescopes (with an incident angle of less than one degree), such as $W1$, is typically less than 2\% of the total geometrical area of the optics. The effective area decreases further when mirror reflectivity is taken into account. Consequently, astronomical X-ray telescopes are developed with coaxially aligned multiple optical shells sharing a common focal plane. These shells are arranged so that the intersection circle of each shell lies on a common sphere, satisfying the sine condition for the entire multi-shell telescope.

The $W1$ design, which is free from spherical aberration and thus provides stigmatic imaging for the on-axis, has coma, astigmatism, and field curvature aberrations, leading to poor imaging performance for off-axis angles. When observing point-like X-ray sources (angular diameter of less than an arcsec) such as black-hole binaries, pulsars, magnetars, and AGN, where the primary objective of the telescope remains to improve the sensitivity of the observation by providing high effective area and good on-axis angular resolution to reduce background noise, $W1$ still remains a good choice. However, when studying extended sources (angular diameter 5-10 arcmin) such as supernovae, galaxies, and galaxy clusters, where in addition to the high effective area, high angular resolution is also required up to a large field of view (FOV) not only to reduce the background noise but also to obtain the geometrical features of the source, $W1$ design suffers due to poor off-axis angular resolution. Yet, being the best realizable design available at the time, $W1$ has been implemented in nearly all astronomical X-ray telescopes, including Chandra \citep{weisskopf2012chandra} and XMM-Newton \citep{lumb2012x}, since the beginning of X-ray astronomy. When an X-ray telescope is developed for survey purposes, where deep sky observations are desired with very long exposures to observe as many faint sources as possible, a large FOV (around 60 arcmin diameter) with high angular resolution over the entire FOV is required. However, the astronomical X-ray survey telescope, eROSITA, developed based on $W1$, has an on-axis half-energy width (HEW) of 15 arcseconds (broadened due to figure errors and diffraction), which increases to approximately 76 arcseconds at a field angle of 30 arcminutes \citep{predehl2010erosita}, primarily due to optical aberrations in the $W1$ design. Therefore, to achieve improved HEW at larger field angles, a better optical design than $W1$ is required. For some special requirements, such as observing high-energy transient sources like gamma-ray bursts (GRBs) after receiving alerts from high-energy all-sky monitoring satellites such as Daksha \citep{bhalerao2024science, bhalerao2024daksha}, a high angular resolution and a wide field X-ray telescope is preferred. There again, $W1$ is not very efficient. The X-ray telescope for solar applications, where high-resolution imaging is required over a large FOV with a diameter of one degree, a better option than $W1$ is desired. For example, the Hinode X-ray Telescope (XRT), which employs a $W1$ optical design, has an on-axis RMS image diameter of 1 arcsec, which degrades to around 4.6 arcsec at a field angle of 15 arcmin \citep{golub2007x}.

The Wolter-Schwarzschild (WS) optics is free from spherical aberration, similar to $W1$. However, since $W1$ optics does not satisfy the abbe sine condition, its principal surface deviates from a perfect sphere, leading to coma aberration. $WS$ optics fully satisfies the abbe sine condition, ensuring its principal surface is spherical and thus eliminating coma aberration \citep{wolter1952verallgemeinerte, saha1987general, wolter1971bildfehlerabschatzung}. Hence, in addition to on-axis stigmatic imaging, $WS$ provides better off-axis imaging performance than $W1$. The off-axis angular resolution can be further improved in both designs by utilizing a curved focal surface to minimize the field curvature effect. Since the implementation of a curved focal detector has practical limitations, multiple flat detectors are arranged in a non-planar shape to approximate a curved focal surface, and it has been implemented in Chandra \citep{garmire2003advanced} and XMM-Newton \citep{turner2001european}. The differences in the optical performance of $W1$ and $WS$ become significant only when the field curvature effect is eliminated. Additionally, the performance difference further decreases as the grazing incidence angle is reduced or the mirror length is increased \citep{chase1973wolter}. The next generation of X-ray telescopes, such as Lynx, is planned to be developed based on $WS$ optical design \citep{schwartz2019lynx}.

Various authors have presented a detailed aberration analysis of $W1$ and $WS$ in the past \citep{mangus1969optical, werner1977imaging, saha1986transverse, shealy1990formula}. So far, exploring an optical design that provides better off-axis angular resolution than $WS$, while retaining on-axis stigmatic imaging performance, does not seem feasible. \cite{saha2022optical} demonstrates that by utilizing polynomials to define optical prescriptions, optimizing off-axis angular resolution while maintaining on-axis stigmatic imaging converges to the performance of $WS$. 

While $W1$ and $WS$ are designed to provide on-axis stigmatic imaging, practical limitations such as figure errors, diffraction, and detector pixel size degrade on-axis angular resolution. Hence, the on-axis resolution may be compromised up to the practically achievable limit in order to improve off-axis resolution. For wide FOV telescopes, such as solar or astronomical survey telescopes, where off-axis resolution is of greater importance than on-axis, the on-axis resolution may be further compromised. For such telescopes, an optical design that efficiently trades off on-axis resolution with off-axis resolution to improve the overall resolution across the entire FOV is more suitable than $W1$ and $WS$. Therefore, for wide FOV telescopes, imaging performance should be evaluated using the area-weighted average angular resolution (AWAAR) over the entire operational FOV.

One such optical design, the Hyperboloid-Hyperboloid (HH) configuration, which improves its off-axis resolution by compromising axial resolution, provides a better AWAAR compared to $W1$ and $WS$ for wide-field telescopes \citep{harvey2001grazing,2005SPIE.5867..114H, nariai1987geometrical, nariai1988geometric}. It has been used in the SXI on GOES \citep{catura2000performance,2007SPIE.6689E..0IH}. $HH$ design can also be utilized to reduce coma aberration compared to $W1$ \citep{thompson2000systems, nariai1987geometrical}. Another approach to improve AWAAR for wide-field telescopes is to explore new optical prescriptions by utilizing polynomials, which provide improved resolution at far-field angles by compromising axial resolution \citep{werner1977imaging, burrows1992optimal,conconi2010wide,elsner2010methods,2014SPIE.9144E..18S,2016ChOpL..14l3401C,saha2022optical}. However, obtaining an optimal polynomial is an iterative process and heavily depends on computational resources.

Here, we introduce a novel optical design, which we name as the Field-angle Optimized ($FO$) design. This design is coma-free and minimizes the PSF size for a given field angle while maintaining an optimal PSF size across other field angles within the FOV. Hence, it provides better AWAAR compared to $W1$, $WS$, and $HH$ when evaluated for either a flat or curved focal surface. The design's improved performance is invariant with telescope geometrical parameters such as focal length, diameter, and mirror length. Section \ref{sec:optical_design} provides a brief overview of the optical designs of $W1$ and $WS$, and the mathematical formulation of $FO$ is also introduced in this section. A comparative analysis of the imaging performance of $FO$ with $W1$, $WS$, and $HH$ is presented in Section \ref{sec:comparative_analysis}. Section \ref{sec:fo_application} discusses the potential applications of $FO$ design and presents an optical design for a solar telescope utilizing $FO$.

\section{Telescope Optical Design} \label{sec:optical_design}

\begin{figure}[!h]
\begin{center}
\includegraphics[width=0.7\linewidth]{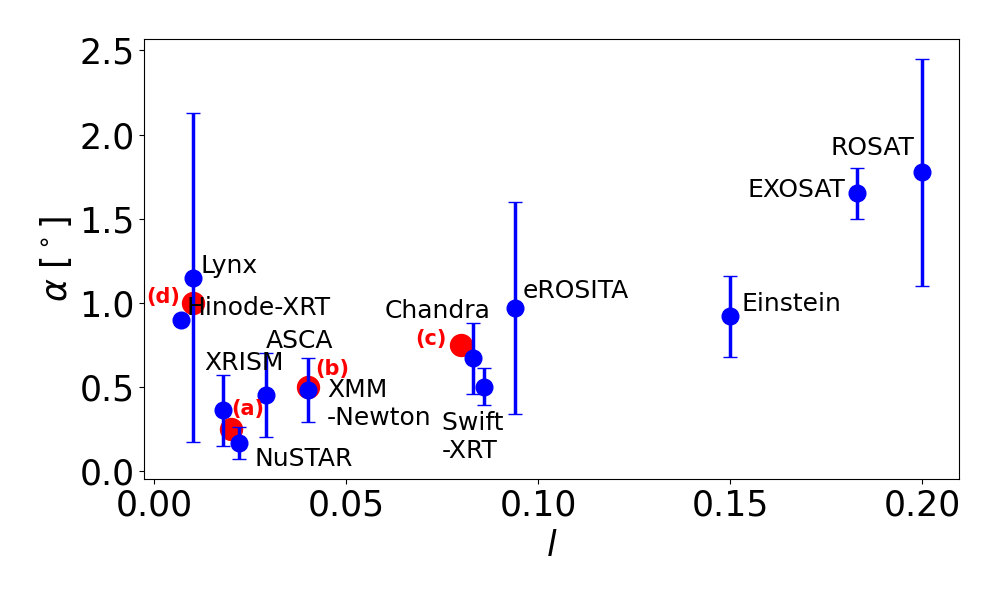}
  \caption{The $\alpha$ and $l$ values for the Lynx \citep{schwartz2019lynx}, Hinode-XRT \citep{golub2007x}, XRISM \citep{iizuka2018ground}, NuSTAR \citep{zhang2009manufacture}, ASCA \citep{serlemitsos1995x}, XMM-Newton \citep{lumb2012x}, Chandra \citep{weisskopf2012chandra}, Swift-XRT \citep{burrows2005swift}, eROSITA \citep{merloni2012erosita}, Einstein \citep{giacconi1979einstein}, EXOSAT \citep{de1981exosat}, and ROSAT \citep{aschenbach1988design} telescopes.}
  \label{fig:telescopes}
\end{center}
\end{figure}

In X-ray telescopes, a single-shell optical system, which consists of two mirrors, can be defined by \textit{$x_0$} (distance between the focal plane and intersection plane (IP) of two mirrors), diameter \textit{D} of the optics at IP,  length of the primary and secondary mirror $L_p$ and $L_s$ (In this work, we consider $L_p=L_s=L$ ), optical prescriptions of both the mirror and the reflectivity property of both the mirror as a function of incident angle and energy of the X-ray photon. The angular resolution of the optics depends only on the ratios of $D/x_0$, $L/x_0$, the mirror's reflectivity, and optical prescriptions. Whereas the effective area of the telescope depends on $D$, $x_0$, $L$, and the mirror's reflectivity, it remains independent of the optical prescription. If we consider the incident angle of the on-axis ray with both the primary and secondary mirrors to be the same as $\alpha$ at the IP, which is required for the optimum effective area of the shell \citep{vanspeybroeck1972design}, and if we further consider that the on-axis rays hitting the primary mirror very close to the IP will focus on the optical axis at a distance \textit{$x_0$} from the IP. Then, $D/x_0$ can be obtained as $D/x_0=2\tan4\alpha$. Hence, to study only the angular resolution of an optical design, a single-shell telescope can be defined by $\alpha$ and $l = L/x_0$. A multishell telescope can be defined as multiple coaxially aligned shells, each with different values of $\alpha$ and $l$.
Figure \ref{fig:telescopes} shows the $\alpha$ and $l$ used in past X-ray telescopes and proposed future telescopes, both solar and astronomical. The vertical bars cover the range of $\alpha$ corresponding to multiple shells present in a telescope, with $\alpha$ increasing from inner shells to outer ones. Each telescope's value of $l$ is the same for all shells. Figure \ref{fig:telescopes} also shows the four red points ((a): $\alpha=0.25^\circ$, $l=0.02$, (b): $\alpha=0.50^\circ$, $l=0.04$, (c): $\alpha=0.75^\circ$, $l=0.08$, (d): $\alpha=1^\circ$, $l=0.01$) which are used for the comparative analysis of $FO$ with $W1$ and $WS$ in the later sections.

In this paper, we use a coordinate system with respect to the telescope such that the axial direction of the telescope aligns with the x-axis. The focal plane lies in the yz-plane with the axial position at x=0 in an ideal configuration. The focal plane can displace axially to minimize field curvature at a particular field angle. Since the telescope is axially symmetric, a point source position can be defined by $\theta$ and $\phi$, where $\theta$ is the angle of the source with respect to the x-axis (defined as the field angle), and $\phi$ represents the azimuthal angle measured from the y-axis towards the +z direction.

\subsection{Wolter-1 and Wolter–Schwarzschild} \label{sec:w_ws}

$W1$ and $WS$ are both free from spherical aberration, ensuring stigmatic imaging in the axial direction; however, their off-axis imaging suffers from various other optical aberrations. While $WS$ is corrected for coma, $W1$ is not, which allows $WS$ to outperform $W1$. This can be demonstrated by Equations \ref{Eq:w_empirical} and \ref{Eq:ws_empirical}. 

\begin{equation}\label{Eq:w_empirical}
    \sigma_{\text{rms}} = 0.2 \frac{\tan^2 \theta}{\tan \alpha} l + 4 \tan \theta \tan^2 \alpha \hspace{1cm} (W1)
\end{equation}

\begin{equation}\label{Eq:ws_empirical}
    \sigma_{\text{rms}} = 0.27 \frac{\tan^2 \theta}{\tan \alpha} l \hspace{1cm} (WS)
\end{equation}

Equation  \ref{Eq:w_empirical} and \ref{Eq:ws_empirical} are the empirical equations first introduced by \cite{vanspeybroeck1972design} and \cite{chase1973wolter} to define the rms image radius of $W1$ and $WS$ as a function of the field angle ($\theta$) and other telescope parameters such as $\alpha$ and $l$ (analytically derived by \citealp{shealy1990formula, nariai1988geometric}). Here, $\sigma_{rms}$ is evaluated at the curved focal surface, and the length of the secondary mirror is assumed to be the same as that of the primary mirror. These equations are valid for $0.5^\circ \leq \alpha \leq 3.5^\circ$, $0.035 \leq l \leq 0.176$, and $\theta \leq 0.5^\circ$. Considering these equations, $WS$ will outperform $W1$ only when $\tan\theta < 57.14\tan^3\alpha/l$. Thus, for a small value of $\alpha$ and a large value of $l$, $WS$ outperforms $W1$ only in a very narrow central region of the FOV. For example, for a telescope with a FOV of $\pm 30$ arcmin, $\alpha = 0.5^\circ$, and $l = 0.04$ (case (b) in Figure \ref{fig:telescopes}), $WS$ outperforms $W1$ only when the field angle is less than 3.26 arcmin. Hence, in such cases, the difference between the performance of $W1$ and $WS$ is negligible.

Equations \ref{Eq:w_empirical} and \ref{Eq:ws_empirical} also show that decreasing \( l \) will improve the telescope's angular resolution. This contradicts the effective area requirement, necessitating a large value of \( l \) to enhance the telescope's effective area. The solution is to utilize a multishell telescope with many shells, each with a smaller value of \( l \). This approach limits the degradation of angular resolution while achieving a high effective area.

\subsection{Field-angle Optimized Design} \label{sec:nd}\label{sec:NewOptcalDesign}
It is well established that the off-axis angular resolution of an X-ray telescope cannot be significantly improved beyond the $WS$ while maintaining on-axis stigmatic imaging \citep{saha2022optical}. Further enhancement of off-axis angular resolution can be achieved by relaxing the on-axis stigmatic imaging constraint, which leads to degradation of on-axis angular resolution. In the $WS$ design, angular resolution is maximized for the on-axis, and the design is free from coma aberration. In contrast, for the $FO$, we maximize angular resolution at a targeted field angle rather than at the on-axis position and ensure that coma aberration is also eliminated. Therefore, we approached the $FO$ with the following constraints:

\begin{description}
  \item[Constraint-1:] When it is optimized to achieve minimum rms image radius at a targeted field angle $\theta_t$, all meridional parallel rays (both at $\phi=0^\circ$ and $\phi=180^\circ$) at $\theta_t$ should converge to a single point.
  \item[Constraint-2:] Optical design should be free from coma aberration.
\end{description}

As a result of these constraints, $FO$ is already fully constrained, which prevents further control over other optical parameters, such as spherical aberration. Therefore, the minimization of the rms image radius at $\theta_t$ is achieved by compromising the on-axis angular resolution of the telescope. Note that since the design is free from coma aberration when it is optimized for $\theta_t=0$, this design should provide imaging performance similar to $WS$.

We found that, mathematically, it is not feasible to define the primary mirror \( f(x_1) \) and secondary mirror \( g(x_2) \) using single continuous functions that meet both constraints mentioned above. Instead, we can derive prescriptions with multiple functions, allowing the mirror to be divided into segments, each represented by different mathematical forms. The complete mirror profile created by these segments is continuous and differentiable. This is achieved by imposing constraints on the boundaries of each segment to ensure continuity at junctions and matching slopes between adjacent segments. Note that while the mathematical definition of the curve is composed of multiple curves, physically, they are single, continuous, and differentiable curves. From a fabrication point of view, they are as realistic as $W1$ and $WS$ or any other curve defined by a single mathematical function.

\begin{figure}[!h]
    \centering
    \begin{subfigure}{0.65\textwidth}
        \centering
    \includegraphics[clip, trim=60 0 60 0, width=\textwidth]{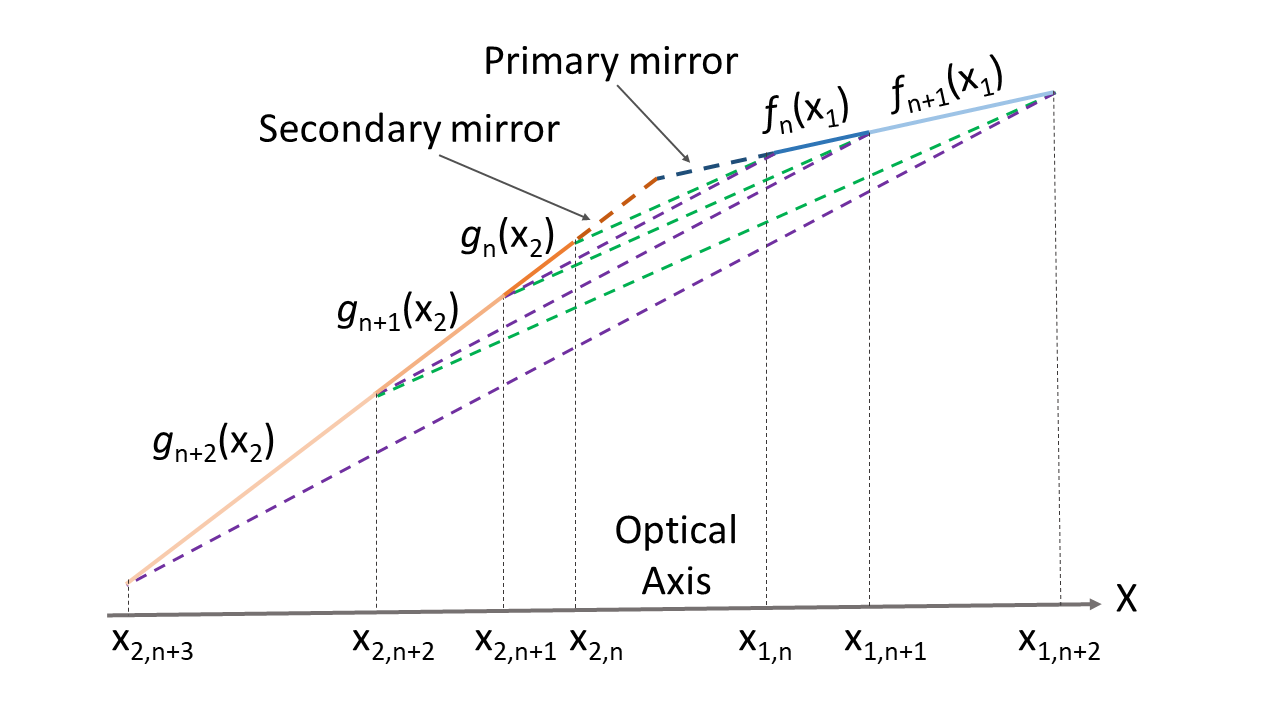}
        \caption{}
        \label{fig:prescription_01}
    \end{subfigure}
    \begin{subfigure}{0.8\textwidth}
        \centering
    \includegraphics[clip, trim=60 0 60 0,width=\textwidth]{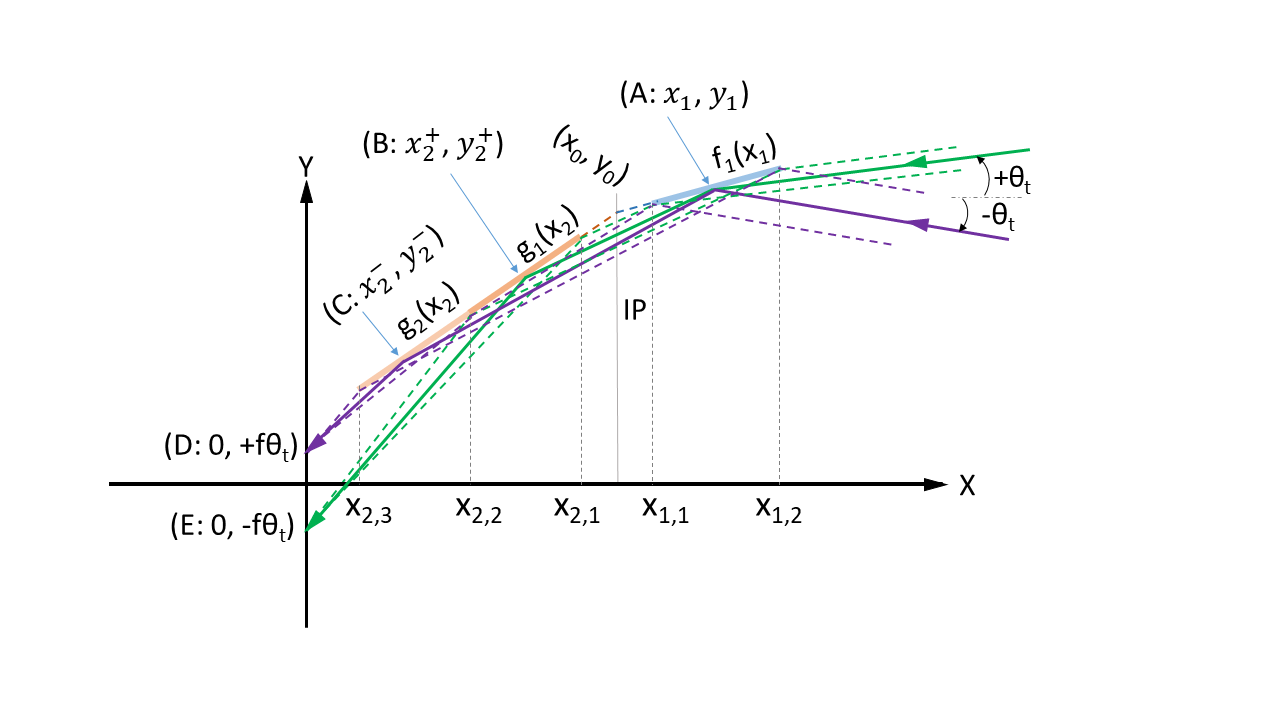}
        \caption{}
        \label{fig:prescription_02}
    \end{subfigure}
    \caption{(a) The primary and secondary mirrors consist of multiple segments that are continuous at the junctions, representing the surface profile for $FO$, (b) The ray trajectory diagram shows two meridional rays: one at $\phi = 0^\circ$ (represented by the solid green line) and one at $\phi = 180^\circ$ (represented by the solid purple line and mirror-imaged along the x-axis) for $FO$ at a field angle of $\theta_t$. Both rays pass through the same point on a segment of the first mirror and are reflected from two consecutive segments of the secondary mirror.}
    \label{fig:prescription}
\end{figure}

Let's say we have two functions $f(x_1)$  and $g(x_2)$, representing the primary and secondary mirrors with multiple functions, as shown in Figure \ref{fig:prescription_01}. These functions can be described as follows. 

\begin{equation}
f(x_1)=\sum\limits_{n=1}^{N} \big( f_n(x_1) : x_{1,n} < x_1 < x_{1,{n+1}}\big)
\end{equation}

\begin{equation}
g(x_2)=\sum\limits_{n=1}^{N} \big(g_n(x_2) : x_{2,{n+1}} < x_2 < x_{2,{n}}\big)
\end{equation}

Here, the boundary conditions are such that, $f_n(x_{1,n+1})=f_{n+1}(x_{1,n+1})$, and $f^\prime_n(x_{1,n+1})=f^\prime_{n+1}(x_{1,n+1})$ with $(^\prime)$ denotes the derivative of the function and the same conditions apply to $g(x_2)$. As we explain later, if $g_{n}(x_2)$ is known, by applying constraint-1 and constraint-2, we can obtain $f_n(x_1)$ and $g_{n+1}(x_2)$. Once the $g_{n+1}(x_2)$ is known, again by applying both the constraints, $f_{n+1}(x_1)$ and $g_{n+2}(x_2)$ can be solved. By sequentially applying the same technique, the remaining segments can be added to both curves to construct them up to the desired mirror lengths. 

To satisfy constraint-1, all meridional rays at $\phi=0^\circ$ and $180^\circ$, with a targeted field angle $\theta_t$, should reach the focal plane at the same position, $y=-f\theta_t$, where $f$ is the focal length of the telescope. To simplify the mathematical process, rays at $\phi=180^\circ$ can be mirrored along the x-axis. Thus, constraint-1 can be redefined: all parallel rays at $\theta=\theta_t$ and $\phi=0^\circ$ should meet the focal plane at $y=-f\theta_t$, while all parallel rays at $\theta=-\theta_t$ and $\phi=0^\circ$ should meet the focal plane at $y=f\theta_t$.

In Figure \ref{fig:prescription_02}, we describe the meridional ray tracing through three mirror segments $f_1(x_1): x_{1,1} < x_1 < x_{1,2}$, $g_1(x_2): x_{2,2} < x_2 < x_{2,1}$ and $g_2(x_2):x_{2,3} < x_2 < x_{2,2}$, and we show how to obtain $f_1(x_1)$ and $g_2(x_2)$, if the $g_1(x_2)$ is known. Equations \ref{Eq:rtrace_01} and \ref{Eq:rtrace_02}, which satisfy the constraint-1 for $\phi=0^\circ$, state that a meridional ray (represented by a solid green ray in Figure \ref{fig:prescription_02}) at $\theta=+\theta_t$ reaches the focal plane at $y_d=-f\theta_t$. Equations \ref{Eq:rtrace_03} and \ref{Eq:rtrace_04} describe the constraint-1 for $\phi=180^\circ$, stating that the meridional ray (represented by a solid purple ray in Figure \ref{fig:prescription_02}) at $\theta=-\theta_t$ reaches the focal plane at $y_d=f\theta_t$. Figure \ref{fig:prescription_02} also shows the meridional rays passing through the boundaries of the $f_1(x_1)$ at $\theta=\theta_t$ and $\theta=-\theta_t$, represented by dashed green and purple rays, respectively.

\begin{equation}\label{Eq:rtrace_01}
{y_1-y^+_{2}=(x_1-x^+_{2})\tan(2\beta_1-\theta_t)}
\end{equation}

\begin{equation}\label{Eq:rtrace_02}
{y^+_{2}+f\theta_t=x^+_{2}\tan(2\beta^+_{2}-2\beta_1+\theta_t)}
\end{equation}

\begin{equation}\label{Eq:rtrace_03}
{y_1-y^-_{2}=(x_1-x^-_{2})\tan(2\beta_1+\theta_t)}
\end{equation}

\begin{equation}\label{Eq:rtrace_04}
y^-_{2}-f\theta_t=x^-_{2}\tan(2\beta^-_{2}-2\beta_1-\theta_t)
\end{equation}

Here, $f$ represents the focal length of the telescope, and $\beta_1$, $\beta^+_2$, and $\beta^-_2$ are the slope angles of the curves at points $A$, $B$, and $C$, respectively. The segment $f_1(x_1)$ can be solved by utilizing Equations \ref{Eq:rtrace_01} and \ref{Eq:rtrace_02}, provided $g_1(x_2)$ is already known. Once the  $f_1(x_1)$ is solved, $g_2(x_2)$ can be solved by utilizing the Equations \ref{Eq:rtrace_03} and \ref{Eq:rtrace_04}.

To satisfy constraint-2, the focal length value should be $f=\sqrt{x_0^2+y_0^2}$, where $y_0$ represents the radius of the intersection circle of the two mirrors at the IP. Note that the value of the focal length deviating from this value will introduce coma aberration. This constraint ensures that the $y$ position on the focal plane of meridional rays at $\theta_t$ originating from the IP should be the same as the $y$ position on the focal plane of sagittal rays originating from the IP. This constraint has also been explained by \cite{werner1977imaging}. Lack of this constraint in $W1$ caused the introduction of coma aberration \citep{wolter1971bildfehlerabschatzung, nariai1987geometrical}.

The boundary values of $f_1(x_1)$,  $g_1(x_2)$ and $g_2(x_2)$ can be obtained as following. When, $x_1$ is considered as $x_{1,1}$, $x^+_2$ and $x^-_2$,  will represent $x_{2,1}$ and $x_{2,2}$, respectively (see Figure \ref{fig:prescription}). And when $x_1$ is considered as $x_{1,2}$, $x^+_2$ and $x^-_2$,  will represent $x_{2,2}$ and $x_{2,3}$, respectively. Hence, the overall philosophy of the design goes in such a way that all rays incident on a segment of the primary mirror, say $f_1(x_1)$, at an angle of $\theta_t$ will be received by segment $g_1(x_2)$ of the secondary mirror. Meanwhile, all rays incident on $f_1(x_1)$ at an angle of $-\theta_t$ will be received by the adjacent segment $g_2(x_2)$ of the secondary mirror.  Additionally, all rays incident on $f_2(x_1)$ (next adjacent segment to $f_1(x_1)$) at an angle of $\theta_t$ will also be received by $g_2(x_2)$. This can be seen in Figure \ref{fig:prescription_01}, where the purple rays joining the two mirrors and passing through the segments' junctions are the rays that hit the primary mirror at an incident angle $-\theta_t$, and the green rays are those that hit the primary mirror at an incident angle $\theta_t$.

Since we have considered an adjoint configuration, in Figure \ref{fig:prescription_02}, we start $x_1$ very close to $x_0$, hence initial segments, $f_1(x_1)$, $g_1(x_2)$ and $g_2(x_2)$, are very close to IP and lengths of these segments are extremely small compared to total mirror length. We have mentioned above that to completely construct both mirrors, the mathematical form of the first segment in the secondary mirror $g_1(x_2)$ should be known in advance. One can choose any mathematical form to define this function, as constraints-1 and 2 are independent of the choice of this function. However, the choice of this function should be such that it should be continuous and differential with respect to the adjacent segment $g_2(x_2)$ at the junction. Once this requirement is met at the first junction, it will automatically make sure that all the remaining segments are continuous and differential at the junctions.

In our study, we have obtained $g_1(x_2)$ in such a way that any ray at a field angle of $\theta$ ($-\theta_t<\theta<\theta_t$), passing through the primary mirror at $x_{1,1}$, after getting reflected from $g_1(x_2)$, reach the focal plan at $-f\theta$. $g_1(x_2)$ obtained by this method ensures that the function is continuous and differentiable at the junction with respect to the next adjacent segments. In this method, as it is considered that $x_{1,1}$ tends to $x_0$, on-axis rays through IP will get focused at the center of the focal plane. However, on-axial rays striking the primary mirror other than IP will not get focused on the center and hence contribute to spherical aberration. 

The numerical approach for implementing the method described above to obtain the prescription of the $FO$ mirror is detailed in Appendix \ref{secA1}.

\section{Comparative Analysis of Optical Designs} \label{sec:comparative_analysis}
In this section, we compare the imaging performance of $FO$ with $W1$, $WS$, and $HH$ only for single-shell telescopes. The imaging performance of $W1$, $WS$, and $FO$ is evaluated using ray tracing with DarsakX software \citep{tiwari2024darsakx}. To begin the comparison, the criteria for evaluating the imaging quality of a design are defined in the following subsection.

\subsection{Imaging quality criteria}
When evaluating the image quality within the entire operational field of view, the imaging performance of an optical design can be defined by how finely the area within the operational field of view is sampled by the telescope. Therefore, the average size of the PSF, which is used for sampling and should be minimized for better optical design, can be defined as $\pi \bar{\sigma}^2$, where $\bar{\sigma}$ represents the `area-weighted average
geometrical rms image radius', estimated as follows. 

\begin{equation} \label{EQ:AWAAR}
\overline{\sigma}(\theta_{max})=\frac{\int_{0}^{\theta_{max}}2\pi\theta\sigma_{rms}(\theta) d\theta}{\pi \theta_{max}^2}
\end{equation}

Here, $\sigma_{rms}(\theta)$ represents the geometrical rms image radius of an optical design at a field angle $\theta$, and $\theta_{max}$ represents the radius of the operational FOV within which the imaging performance of the telescope needs to be evaluated. Similar criteria have been used to evaluate the imaging performance of a wide FOV imaging telescopes by \cite{burrows1992optimal}, \cite{harvey2001grazing}, \cite{elsner2010methods}, and \cite{conconi2010wide}.

\subsubsection{At flat focal surface}
When $\bar{\sigma}$ for an optical design is evaluated at a flat focal surface, its value can be further minimized by adjusting the axial position of the focal plane. $\sigma_{rms}$ depends not only on $\theta$ but also on the focal plane axial position ($\Delta x$, which is the axial displacement of the focal plane from the focal point of the axial rays). By displacing the focal plane, $\sigma_{rms}$ can be minimized for a particular field angle $\theta$ by eliminating the field curvature effect at this field angle. However, this adjustment may degrade $\sigma_{rms}$ at other field angles, yet still resulting in a reduced area-weighted average $\bar{\sigma}$ over the entire FOV. Hence, in the case of a flat focal surface, Equation \ref{EQ:AWAAR} is modified as follows.

\begin{equation} \label{EQ:AWAAR_01}
\overline{\sigma}(\theta_{max}, \Delta x)=\frac{\int_{0}^{\theta_{max}}2\pi\theta\sigma_{rms}(\theta,\Delta x) d\theta}{\pi \theta_{max}^2}
\end{equation}

In case of $W1$ and $WS$, the  $\bar{\sigma}(\theta_{max}, \Delta x)$ gets minimize at a non-zero $\Delta x$, and its value depends on the $\theta_{max}$ as well. In case of $FO$, $\sigma_{rms}$ depends not only $\theta$ and $\Delta x$ but also depends on $\theta_t$. Because each $\theta_t$ provides a different set of prescriptions of the optical mirror so that $\sigma_{rms}$ is minimized at $\theta_t$. However, in the case of $FO$, we found that the minimum value of $\bar{\sigma}$ is always obtained when $\Delta x=0$, hence we can say for $FO$ at the flat focal surface, $\bar{\sigma}$ can be evaluated by following equation.

\begin{equation} \label{EQ:AWAAR_02}
\overline{\sigma}(\theta_{max}, \theta_t)=\frac{\int_{0}^{\theta_{max}}2\pi\theta\sigma_{rms}(\theta,\theta_t) d\theta}{\pi \theta_{max}^2}
\end{equation}

To compare various optical designs, a minimum possible value of $\pi \bar{\sigma}^2$ needs to be evaluated for each design for a given operational field of view ($\theta\leq \theta_{max}$). Hence, for $W1$ and $WS$ design for a given $\theta_{max}$, $\bar{\sigma}$ need to be minimized over $\Delta x$. Similarly, for $FO$, $\bar{\sigma}$ need to be minimized over $\theta_t$.

\subsubsection{At curved focal surface}
At the curved focal surface, field curvature is already minimized for all field angles, and hence $\sigma_{rms}$ cannot be further minimized at any field angle by displacing the curved focal surface. Hence, for $W1$ and $WS$, $\bar{\sigma}(\theta_{max})$ can be defined by Equation \ref{EQ:AWAAR}. But in the case of $FO$ at the curved focal surface, the variation of $\sigma_{rms}$ with field angle can be controlled with $\theta_t$. Hence, again the Equation \ref{EQ:AWAAR_02} will be applicable in this case.

\subsection{Imaging quality comparison}

\begin{figure}[!h]
    \centering
    \begin{subfigure}[b]{0.48\textwidth}
    \includegraphics[width=\textwidth]{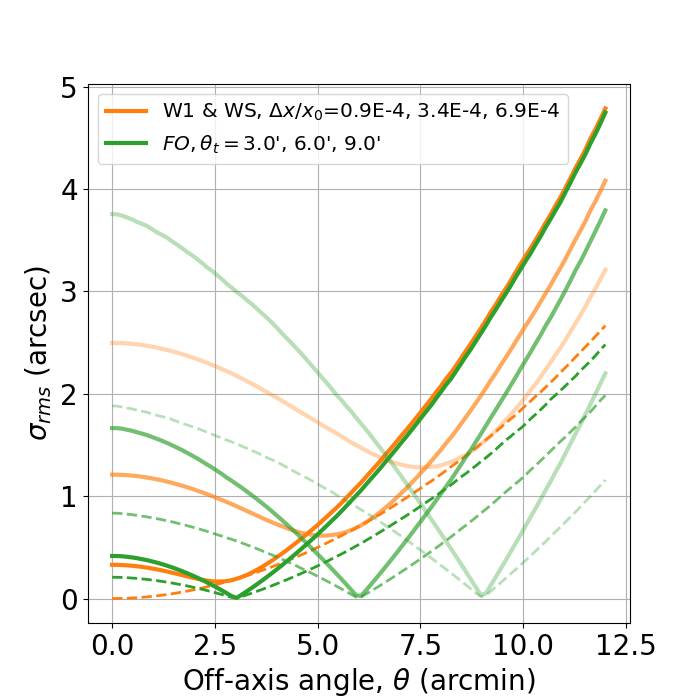}
        \caption{$\alpha=0.25^\circ$, $l=0.02$}
        \label{fig:sigma_rmsa}
    \end{subfigure}
    \hspace{-0.5cm}
    \begin{subfigure}[b]{0.48\textwidth}
        \includegraphics[width=\textwidth]{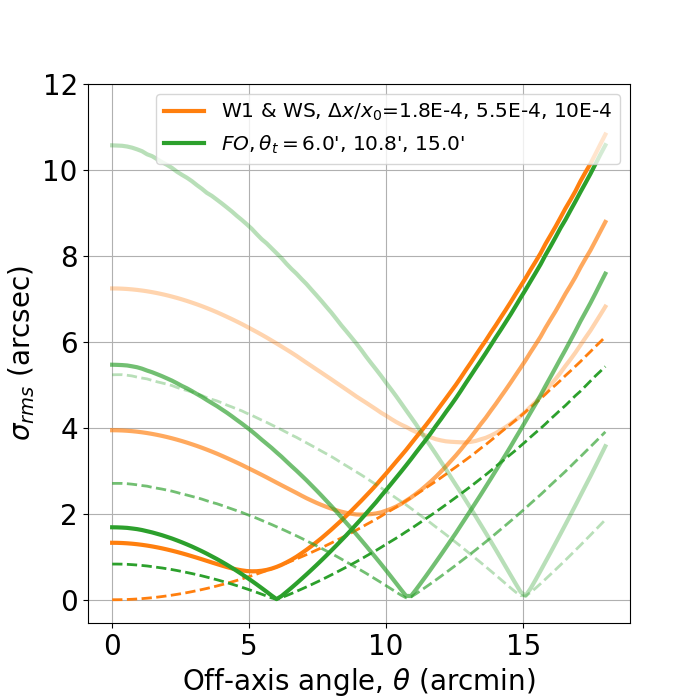}
        \caption{$\alpha=0.5^\circ$, $l=0.04$}
        \label{fig:sigma_rmsb}
    \end{subfigure}
    \vspace{0cm}
    \begin{subfigure}[b]{0.48\textwidth}
        \includegraphics[width=\textwidth]{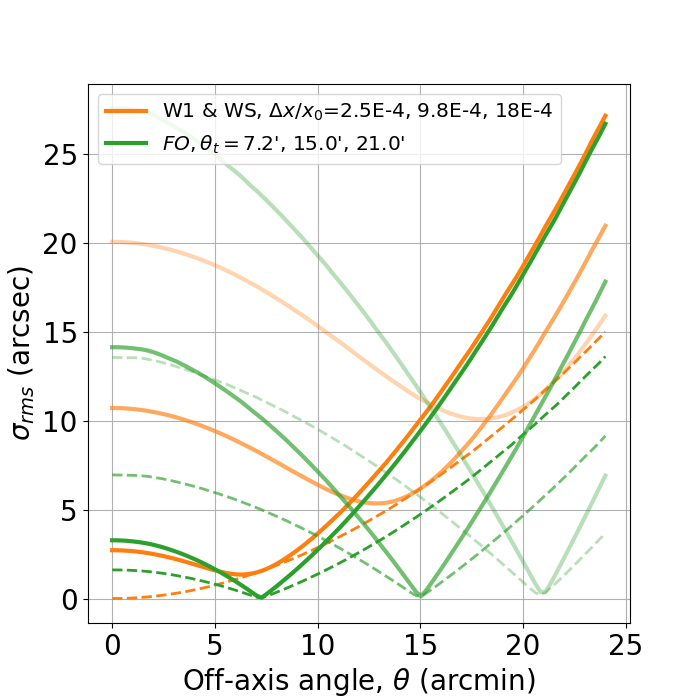}
        \caption{$\alpha=0.75^\circ$, $l=0.08$}
        \label{fig:sigma_rmsc}
    \end{subfigure}
    \hspace{-0.5cm}
    \begin{subfigure}[b]{0.48\textwidth}
        \includegraphics[width=\textwidth]{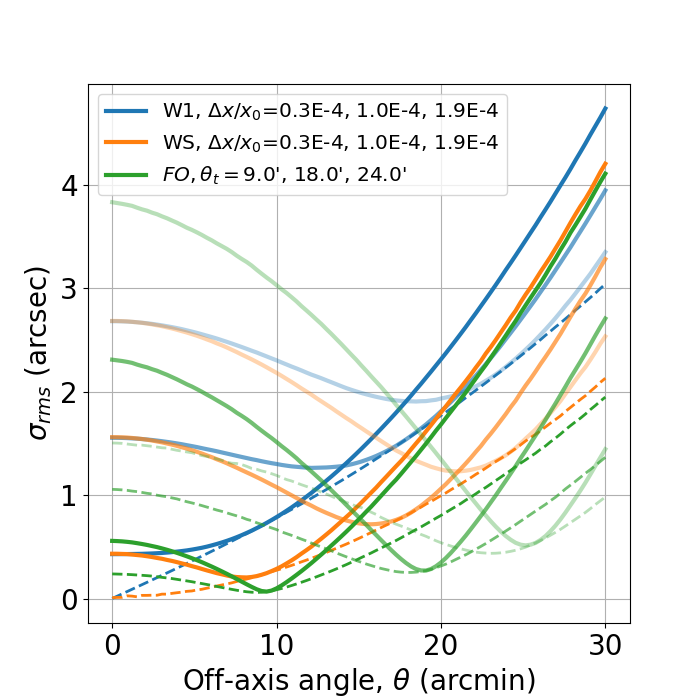}
        \caption{$\alpha=1^\circ$, $l=0.01$}
        \label{fig:sigma_rmsd}
    \end{subfigure}
    \caption{$\sigma_{rms}(\theta)$ is plotted for $W1$, $WS$, and $FO$ for four sets of values of $\alpha$ and $l$. For $W1$ and $WS$, $\sigma_{rms}(\theta)$ is plotted at three axial positions with a flat focal surface and at one position with a curved focal surface. For $FO$, $\sigma_{rms}(\theta)$ is plotted for three values of $\theta_t$, evaluated at both flat and curved focal surfaces. Solid lines represent $\sigma_{rms}(\theta)$ evaluated at a flat focal surface, and dashed lines represent $\sigma_{rms}(\theta)$ evaluated at a curved focal surface. For $W1$ and $WS$ at a flat focal surface, an increment in the displacement of the focal plane ($\Delta x/x_0$) is indicated by increasing the transparency of the curves. Similarly, for $FO$, an increment in $\theta_t$ is represented by increasing the transparency of the curves.}
    \label{fig:sigma_rms}
\end{figure}

The $\sigma_{rms}(\theta)$ for $W1$, $WS$, and $FO$ are shown in Figure \ref{fig:sigma_rms} for four different optical designs (four sets of $\alpha$ and $l$). The values of $\alpha$ and $l$ for these four optical designs, compared to previously flown and future telescope designs, are shown in Figure \ref{fig:telescopes} marked with red color. For $FO$ in all four designs, $\sigma_{rms}(\theta)$ with flat and curved focal surfaces is plotted for three different values of $\theta_t$. For $W1$ and $WS$, $\sigma_{rms}(\theta)$ with a flat focal surface is plotted for three different focal plane axial positions ($\Delta x/x_0$). The focal plane is displaced in such a way that the field curvature is minimized at field angles identical to the values of $\theta_t$ as in $FO$. Additionally, for $W1$ and $WS$, $\sigma_{rms}(\theta)$ is also plotted with a curved focal surface.

In Figures \ref{fig:sigma_rmsa}, \ref{fig:sigma_rmsb}, and \ref{fig:sigma_rmsc}, $\alpha$ is low and $l$ is high compared to Figure \ref{fig:sigma_rmsd}. As a result, the $\sigma_{rms}(\theta)$ for $W1$ and $WS$ in Figures \ref{fig:sigma_rmsa}, \ref{fig:sigma_rmsb}, and \ref{fig:sigma_rmsc} are nearly identical and overlaps. Hence, their curves are combined into a single curve. However, in the case where $\alpha$ is high and $l$ is low, as represented by Figure \ref{fig:sigma_rmsd}, it is clear that $WS$ outperforms $W1$. Such combinations of high $\alpha$ and low $l$ are typically applicable for solar telescopes in very soft X-ray range (energy $<$2keV). It can be seen that in all four cases, whenever the aim is to minimize $\sigma_{rms}$ for a non-zero $\theta_t$, $FO$ consistently provides lower $\sigma_{rms}(\theta_t)$ compared to $W1$ and $WS$. For cases \ref{fig:sigma_rmsa} to \ref{fig:sigma_rmsc}, $\sigma_{rms}(\theta_t)$ for $FO$ is very close to zero. However, in the case where $\alpha$ is high and $l$ is low, represented by \ref{fig:sigma_rmsd}, $\sigma_{rms}(\theta_t)$ has a finite value and it continuously increases with $\theta_t$. It is also noticed that, as $\theta_t$ tends to zero, $FO$ converges to $WS$, as expected. Figure \ref{fig:sigma_rms} concludes that while $WS$ can provide $\sigma_{rms} = 0$ only for on-axis sources, $\sigma_{rms}$ continuously increases with the off-axis angle and cannot be reduced beyond the value achieved at the curved focal surface. On the other hand, $FO$ can achieve $\sigma_{rms} \approx 0$ for on-axis as well as for any desired off-axis angle in cases (a) to (c) (shown for only three off-axis angles in Figure \ref{fig:sigma_rms}). In case (d), although $\sigma_{rms}$ is not close to zero, it is significantly lower than for $WS$. The ability of $FO$ to achieve lower $\sigma_{rms}$ at off-axis angles enables it to achieve higher imaging performance on an average scale, resulting in lower $\overline{\sigma}$ compared to $WS$.

\begin{figure}
\begin{center}
\includegraphics[trim={0 1cm 0 0},clip,width=0.6\linewidth]{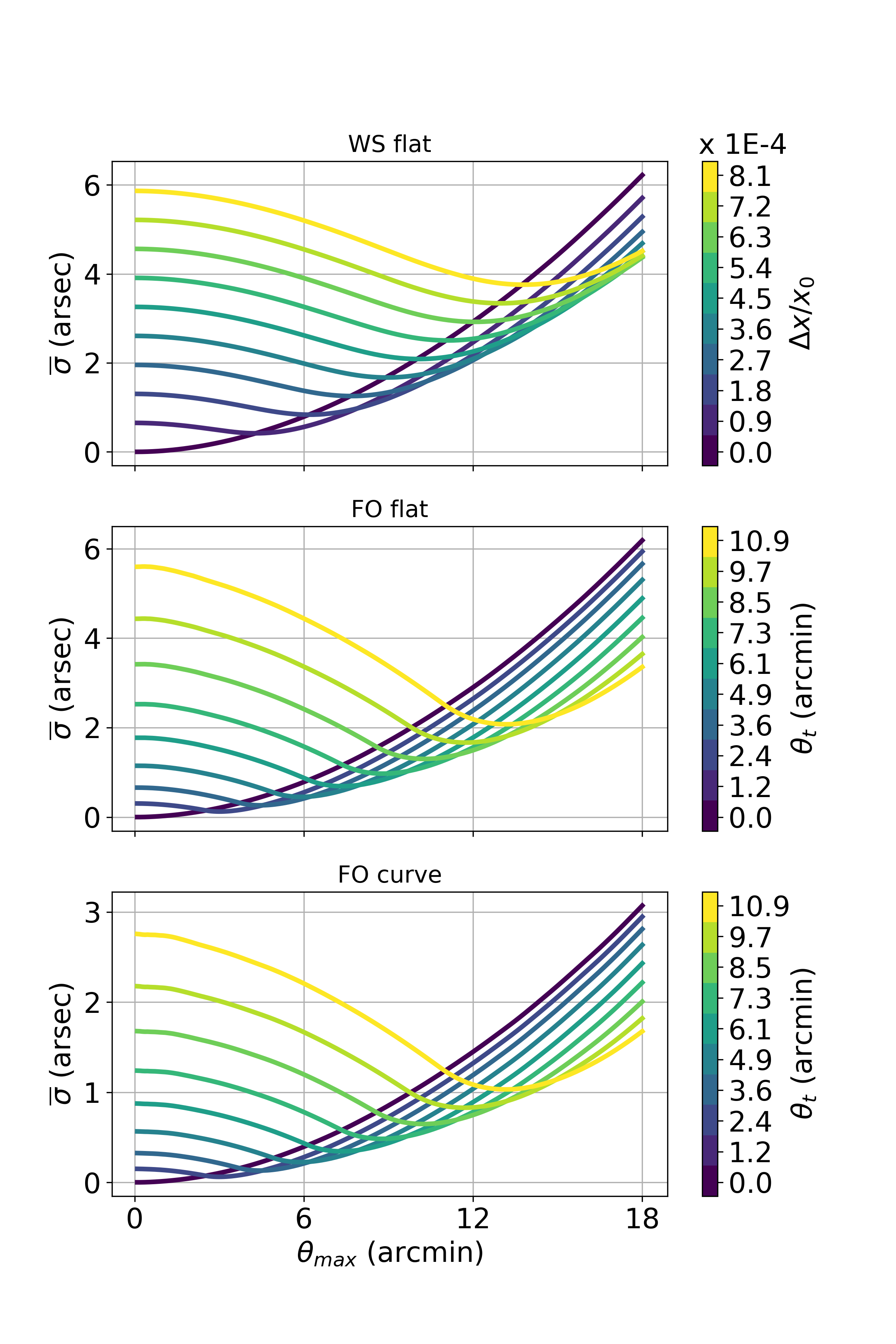}
  \caption{$\overline{\sigma}(\theta_{max})$ for $WS$ on a flat focal surface at various axial locations (top), and for $FO$ on flat (middle) and curved (bottom) focal surfaces with various values of $\theta_t$, for telescope parameters of $\alpha = 0.5^\circ$ and $l = 0.04$.}
  \label{fig:sigma_bar}
\end{center}
\end{figure}

To compare the imaging performance of two optical designs, the minimum possible $\overline{\sigma}$ for each design must be determined. For $W1$ and $WS$, when evaluating $\overline{\sigma}$ for a flat focal surface, its value varies with the axial position of the focal surface. Therefore, the minimum possible $\overline{\sigma}$ is found by determining the optimum axial location of the focal plane. Similarly, for $FO$, both at flat and curved focal surfaces, $\overline{\sigma}$ depends on the off-axis angle ($\theta_t$) at which $\sigma_{rms}$ is minimized by adjusting the mirror prescription. Thus, the minimum possible $\overline{\sigma}$ is obtained by minimizing it with respect to $\theta_t$. In Figure \ref{fig:sigma_bar}, $\overline{\sigma}(\theta_{max},\Delta x/x_0)$ for $WS$ design is shown which is evaluated at the flat focal surface for a telescope with $\alpha=0.5^\circ$ and $l=0.04$ (case-(b) in Figure \ref{fig:telescopes}). For this set of $\alpha$ and $l$, imaging performance of $W1$ and $WS$ are nearly identical, hence we did the comparison of $FO$ with $WS$ only. In the same plot, $\overline{\sigma}(\theta_{max},\theta_t)$ is also shown for $FO$ at flat and curved focal surface. $\overline{\sigma}(\theta_{max},\Delta x/x_0)$ for $WS$ at flat focal surface can be minimized over $\Delta x/x_0$ by choosing a curve at $\theta_{max}$ which provided minimum $\overline{\sigma}$. $\Delta x/x_0$ obtained by this curve will provide the best possible axial location of a flat focal surface leading to the minimum $\overline{\sigma}$ for the given $\theta_{max}$. Similarly, $\overline{\sigma}(\theta_{max}, \theta_t)$ for $FO$ can be minimized over $\theta_t$ for flat and curved focal surface by choosing appropriate curves at a particular $\theta_{max}$. By this approach, the minimum possible $\overline{\sigma}(\theta_{max})$ can be evaluated for $WS$ at flat focal surfaces and for $FO$ at flat and curved focal surfaces. $\overline{\sigma}(\theta_{max})$ for $WS$ at curved detector can be directly evaluated using Equation \ref{EQ:AWAAR} as it doesn't depends on $\Delta x/x_0$ or $\theta_t$. With a similar approach, the minimum possible $\overline{\sigma}(\theta_{\text{max}})$ can be obtained for the other three telescope designs (case-(a), (c), and (d) in Figure \ref{fig:telescopes}).

\begin{figure}
    \centering
    \begin{subfigure}[b]{0.48\textwidth}
    \includegraphics[clip, trim=0 10 0 18,width=\textwidth]{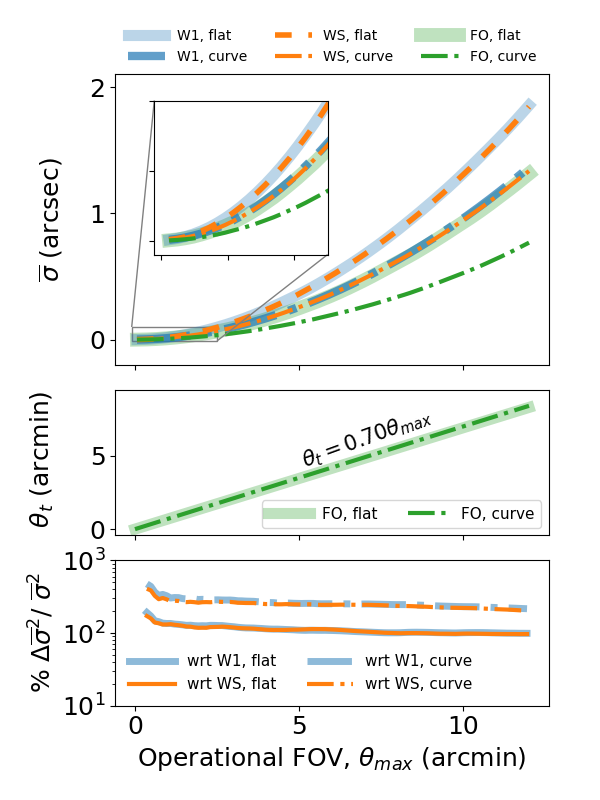}
        \caption{$\alpha=0.25^\circ$, $l=0.02$}
        \label{}
    \end{subfigure}
    \hspace{-0.3cm}
    \begin{subfigure}[b]{0.48\textwidth}
        \includegraphics[clip, trim=0 10 0 18,width=\textwidth]{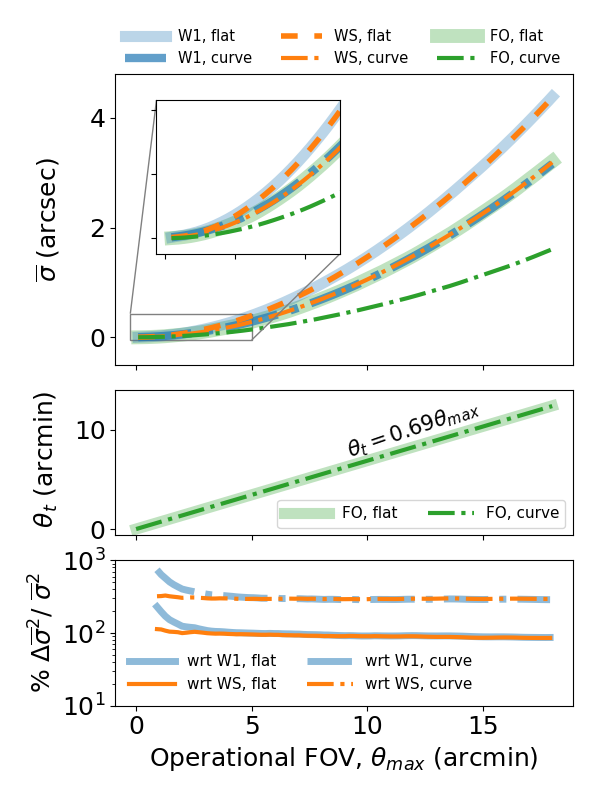}
        \caption{$\alpha=0.5^\circ$, $l=0.04$}
        \label{}
    \end{subfigure}
    \vspace{0cm}
    \begin{subfigure}[b]{0.48\textwidth}
        \includegraphics[clip, trim=0 10 0 18,width=\textwidth]{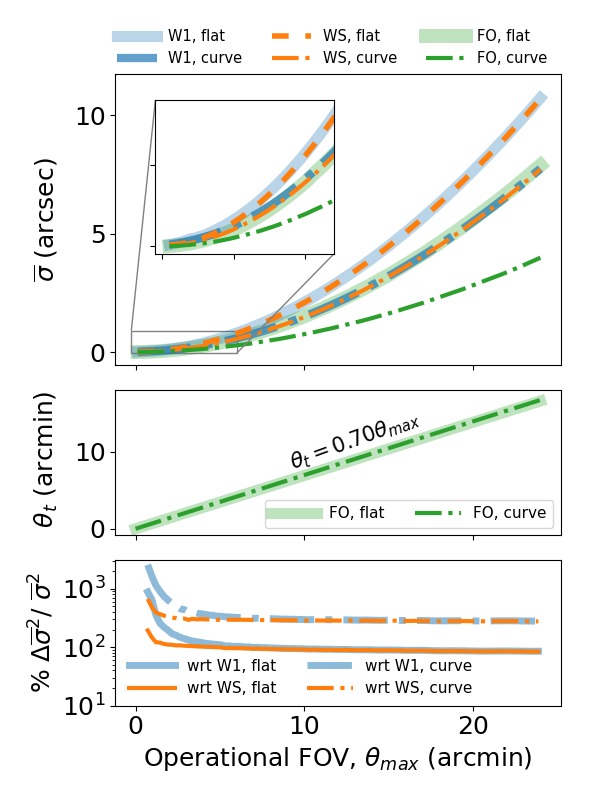}
        \caption{$\alpha=0.75^\circ$, $l=0.08$}
        \label{}
    \end{subfigure}
    \hspace{-0.3cm}
    \begin{subfigure}[b]{0.48\textwidth}
        \includegraphics[clip, trim=0 10 0 18,width=\textwidth]{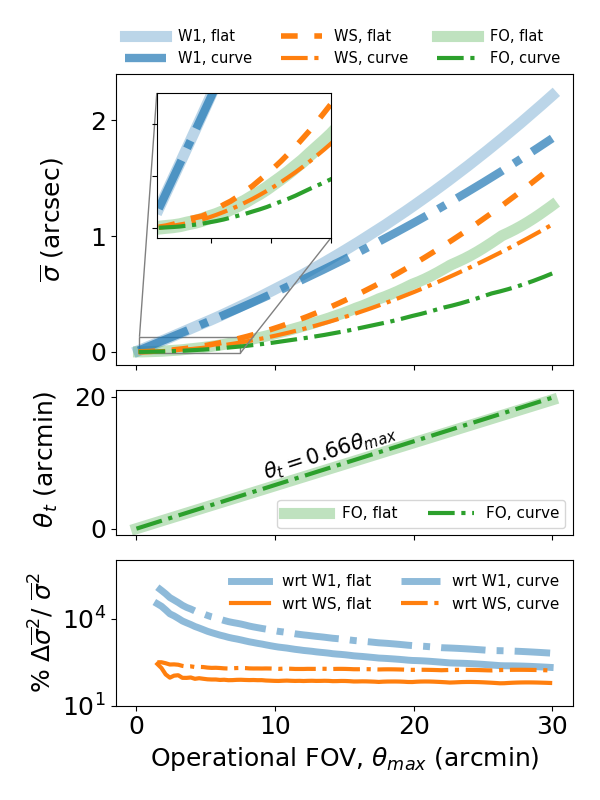}
        \caption{$\alpha=1^\circ$, $l=0.01$}
        \label{}
    \end{subfigure}
    \caption{Four panels (a-d), corresponding to different values of $\alpha$ and $l$, are presented. In each main panel, the top sub-panel shows the comparison of $\overline{\sigma}(\theta_{max})$ for $W1$, $WS$, and $FO$ at flat and curved focal surfaces. The middle sub-panel illustrates the relationship between $\theta_t$ and $\theta_{max}$ when the minimum $\overline{\sigma}$ is obtained for $FO$. The bottom sub-panel displays the percentage increase in the average PSF size ($\pi \overline{\sigma}^2$) of $W1$ and $WS$ compared to $FO$ at flat and curved focal surfaces.}
\label{fig:sigma_bar_theta_max}
\end{figure}

The minimum possible $\overline{\sigma}(\theta_{\text{max}})$ obtained for the $W1$, $WS$, and $FO$ designs, at both flat and curved focal surfaces, for four different telescopes with varying sets of $\alpha$ and $l$, as marked by red points in Figure \ref{fig:telescopes}, is shown in Figure \ref{fig:sigma_bar_theta_max}. It can be seen that $FO$ consistently provides a smaller $\overline{\sigma}$ compared to $W1$ or $WS$ for all four cases, whether at a flat or curved focal surface, for any value of the operational field of view, $\theta_{\text{max}}$. In Figure \ref{fig:sigma_bar_theta_max} for $FO$, we also show the value of the field angle $\theta_t$ at which $\sigma_{\text{rms}}$ is minimized to achieve the minimum $\overline{\sigma}$ within the operational field of view defined by $\theta_{\text{max}}$. It is evident that for $FO$, the ratio of $\theta_t$ to $\theta_{\text{max}}$ is constant and nearly identical for both flat and curved focal surfaces; for all four cases, it is around 0.65 to 0.7. For example, for case-(b), $\theta_t/\theta_{\text{max}} = 0.69$, which means that $FO$ provides the best performance when the rms value of the PSF is minimized at a radial position of $69\%$ of the overall field of view. In Figure \ref{fig:sigma_bar_theta_max}, we also show the percentage increase in the size of the average PSF ($\Delta (\pi {\overline{\sigma}}^2)/\pi {\overline{\sigma}}^2$) for the $W1$ and $WS$ designs compared to $FO$ at both flat and curved focal surfaces. The average PSF size for $W1$ and $WS$ is always larger than that for $FO$ at any value of $\theta_{\text{max}}$. For all four cases, the average PSF size for both $W1$ and $WS$ is significantly higher at a smaller field of view compared to $FO$, and they become constant at larger field of view sizes. If we consider case-(b), as $\theta_{\text{max}}$ increases to 18 arcminutes, the increment in the average PSF size of $W1$ and $WS$ with respect to $FO$ becomes constant, and they are around $87\%$ larger than that of $FO$ at a flat focal surface and around $294\%$ larger at a curved focal surface.

The ratios $\theta_t / \theta_{max}$ and $\Delta \overline{\sigma}^2 / \overline{\sigma}^2$ for all four sets of $\alpha$ and $l$ are tabulated in Table \ref{tab:performance_comparison}. For cases (a), (b), and (c), $\Delta \overline{\sigma}^2 / \overline{\sigma}^2$ at a flat focal surface lies in the range of 84-99$\%$, and at a curved focal surface, it lies in the range of 202-294$\%$ for $W1$ and $WS$, both with respect to $FO$. For case (d), where $\alpha$ is comparatively large and $l$ is small, hence the coma effect is significantly high in the case of $W1$, leading to poorer performance compared to $WS$ and $FO$. For this case, $\Delta \overline{\sigma}^2 / \overline{\sigma}^2$ for $W1$ is quite high, around 208$\%$ at the flat focal surface and 648$\%$ at the curved focal surface with respect to $FO$. Whereas, the difference in $\overline{\sigma}^2$ between $WS$ and $FO$ is reduced compared to other cases (a), (b), and (c). $\Delta \overline{\sigma}^2 / \overline{\sigma}^2$ is around 64$\%$ at the flat focal surface and around 166$\%$ at the curved focal surface.

\begin{table}
    \centering
    \begin{tabular}{@{\extracolsep\fill}p{2.3cm}p{1.5cm}cp{2.3cm}p{2.1cm}}
    \toprule%
        \textbf{Telescope Parameters} & \textbf{Focal Surface} & \textbf{$\theta_t/\theta_{max}$} & \textbf{$\Delta  {\overline{\sigma}}^2$/$ {\overline{\sigma}}^2$ for $W1$ wrt $FO$} & \textbf{$\Delta  {\overline{\sigma}}^2$/$ {\overline{\sigma}}^2$ for $WS$ wrt $FO$} \\
        \midrule
        \multirow{2}{2.3cm}{(a): $\alpha=0.25^\circ$, $l=0.02$} & flat & 0.69 &  99.3$\%$ \newline at $\theta_{max}=12'$ & 97.0$\%$ \newline at $\theta_{max}=12'$\\
         \cmidrule{2-5}
         & curve & 0.70 & 216.9$\%$ \newline at $\theta_{max}=12'$ & 202.9$\%$ \newline at $\theta_{max}=12'$\\
         \midrule
    \multirow{2}{2.3cm}{(b): $\alpha=0.50^\circ$, $l=0.04$} & flat & 0.69 & 86.9$\%$ \newline at $\theta_{max}=18'$ & 85.6$\%$ \newline at $\theta_{max}=18'$\\
             \cmidrule{2-5} 
             & curve & 0.69 &287.9$\%$ \newline at $\theta_{max}=18'$ & 294.3$\%$ \newline at $\theta_{max}=18'$\\
         \midrule
    \multirow{2}{2.3cm}{(c): $\alpha=0.75^\circ$, $l=0.08$} & flat & 0.69 & 85.2$\%$ \newline at $\theta_{max}=24'$&84.3$\%$ \newline at $\theta_{max}=24'$\\
             \cmidrule{2-5} 
             & curve & 0.69 &278.1$\%$ \newline at $\theta_{max}=24'$ & 275.2$\%$ \newline at $\theta_{max}=24'$\\
         \midrule
    \multirow{2}{2.3cm}{(d): $\alpha=1.0^\circ$, $l=0.01$} & flat & 0.65 & 208.4$\%$ \newline at $\theta_{max}=30'$&64.0$\%$ \newline at $\theta_{max}=30'$\\
             \cmidrule{2-5} 
             & curve & 0.66 &648.4$\%$ \newline at $\theta_{max}=30'$ & 166.9$\%$ \newline at $\theta_{max}=30'$\\
         \botrule%
    \end{tabular}
    \caption{Ratio of $\theta_t/\theta_{max}$ for $FO$, and percentage of increment in the average PSF size ($\pi \overline{\sigma}^2$) of $W1$ and $WS$ compared to $FO$ at flat and curved focal surfaces for four different sets of $\alpha$ and $l$.
}
    \label{tab:performance_comparison}
\end{table}

We also compare the $FO$ performance with $HH$. The performance of $HH$ for a telescope design with $\alpha=1.74^\circ$ and $l=0.072$ is discussed by \cite{harvey2001grazing}. There, it has been shown that the improvement in $\Delta \overline{\sigma}^2 / \overline{\sigma}^2$ for $HH$ with respect to $W1$ and $WS$ for a flat focal surface at $\theta_{max}=21'$ is around 85.8\% and 10.6\%, respectively. For the same telescope design, the improvement in $\Delta \overline{\sigma}^2 / \overline{\sigma}^2$ for $FO$ with respect to $W1$ and $WS$ for a flat focal surface at $\theta_{max}=21'$ is around 218\% and 83.3\%, respectively. When compared at a curved focal surface, the performance for $FO$ further improves to 824\% and 280\% with respect to $W1$ and $WS$, respectively. \cite{harvey2001grazing} shows that for this telescope design, $HH$ outperforms $WS$ only when $\theta_{\text{max}}$ is greater than 18$'$. However, $FO$ outperforms $WS$ for any value of $\theta_{max}$ not only for this design but also for other telescope designs represented by a different set of $\alpha$ and $l$ as shown in Figure \ref{fig:telescopes}.

\begin{figure}
    \centering
    \begin{subfigure}{0.45\textwidth}
        \centering
        \includegraphics[width=\textwidth]{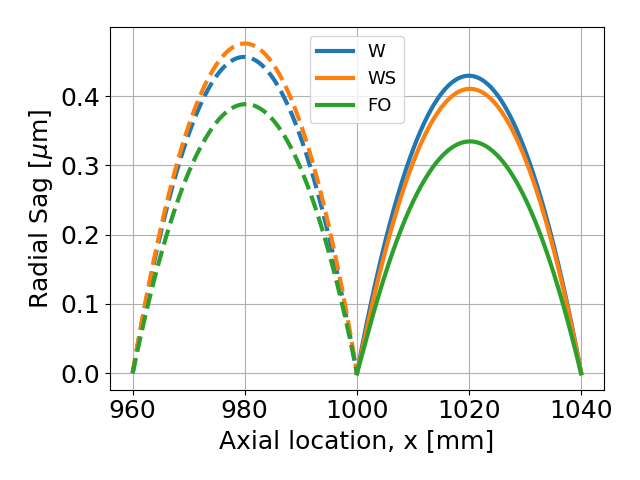}
        \caption{}
        \label{fig:radial_sag}
    \end{subfigure}
    \begin{subfigure}{0.45\textwidth}
        \centering
        \includegraphics[width=\textwidth]{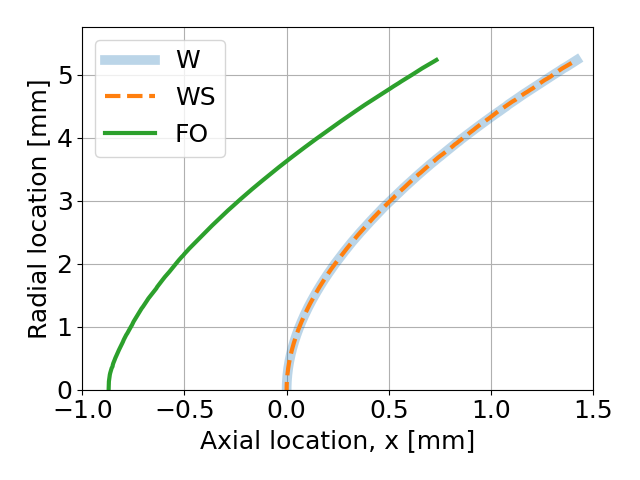}
        \caption{}
        \label{fig:curved_focal_surface}
    \end{subfigure}
    \caption{(a) Sectional view of the radial sag present in the surface profile of the mirrors, (b) sectional view of the optimal curved focal surface to minimize field curvature for a telescope with $\alpha=0.5^\circ$, $l=0.04$, $x_0=1000$ mm, and for $FO$, $\theta_t = 12.42$ arcmin.}
    \label{fig:sag_curved}
\end{figure}

In Figure \ref{fig:radial_sag}, the sectional view ($\phi=0^\circ$) of the radial sag, which represents the surface profile of a mirror after the subtraction of a conical surface, in the primary and secondary mirrors is shown for $W1$, $WS$, and $FO$ for case (b) with $\alpha=0.5^\circ$ and $l=0.04$ (see Figure \ref{fig:telescopes}). The $x_0$ is considered to be 1000 mm, and for $FO$, the $\theta_t=12.42$ arcmin. The surface profile of $FO$ mirrors, although constructed with multiple segments that are continuous and differentiable at junctions, is as feasible as those of $W1$ and $WS$ from a fabrication perspective. For the same telescope parameters, the sectional view of the axially symmetric curved focal surface to minimize the field curvature effect for $W1$, $WS$, and $FO$ is shown in Figure \ref{fig:curved_focal_surface}. The imaging performance of $W1$ and $WS$ is nearly identical; hence, their curved surfaces overlap.

\subsection{PSF comparison}

\begin{figure}[!h]
\begin{center}
\includegraphics[width=0.8\linewidth]{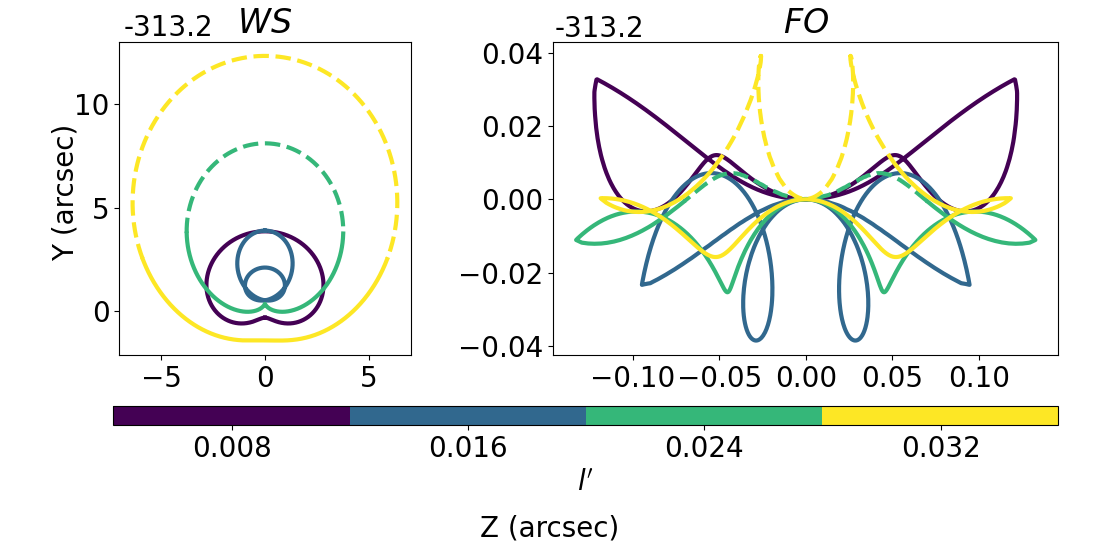}
  \caption{Geometrical ray tracing on the focal plane through various rings on the primary mirror for the optical design of $WS$ and $FO$ in a single-shell telescope with $\alpha = 0.5^\circ$ and $l = 0.04$, with a source located at $\theta = 12.42$ arcmin. For $FO$, $\theta_t = 12.42$ arcmin, and for $WS$, the field curvature is minimized at a field angle of $\theta_t$ by displacing the focal plane axially. $l'$ represents the axial position of the rings on the primary mirror, which are traced to the focal plane and shown as closed loops. Solid lines represent rays traced within the secondary mirror, with a length equal to the primary mirror, while dashed lines represent rays with lengths greater than the primary mirror.}
  \label{fig:psf_comparision}
\end{center}
\end{figure}

In Figure \ref{fig:psf_comparision}, we present the geometrical ray trace onto the focal surface from a point source at infinity, with a field angle of $\theta = 12.42$ arcmin and $\phi = 0^\circ$, for a single-shell telescope with the optical designs of $WS$ and $FO$, having $\alpha = 0.5^\circ$ and $l = 0.04$. Instead of illuminating the primary mirror entirely with parallel rays, we traced the rays through various rings on the primary mirror at different axial positions. Therefore, Figure \ref{fig:psf_comparision} does not present a complete PSF; rather, it shows the tracing of rays through the various rings on the primary mirror to the focal surface. For $FO$, optical prescriptions are also obtained for $\theta_t=12.42$ arcmin. For $WS$, the PSF is estimated over a flat focal surface which is displaced from the origin to minimize the field curvature at $\theta=12.42$ arcmin, and for $FO$, this flat surface is placed at the origin. PSF is represented by four different curves, each represented by a different color. These colors represent the axial position ($l'=(x_P-x_0)/x_0$, where $x_P$ is the axial location of the ring on the primary mirror) of a ring on the primary mirror, from which rays incident at $\theta=12.42$ arcmin are traced to the focal surface and represented by closed color loops. The solid lines represent the rays received by the secondary mirror with a length equal to that of the primary mirror ($l=0.04$). The dashed lines represent the rays received by the secondary mirror in the region where the length of the mirror is greater than that of the primary mirror ($l>0.04$).

The PSF by $FO$ has a butterfly-like structure because of the first constraint defined for $FO$ in Section \ref{sec:NewOptcalDesign}, that the meridional rays at $\phi=0^\circ$ and $180^\circ$ originating from a source at $\theta_t$ will be focused at a single position. Hence, for $FO$, at $z=0$ arcsec, all rays are focused at a single position. The wings are present because this constraint was applied only for $\phi=0^\circ$ and $180^\circ$, not for other values of $\phi$.

\section{$FO$ Application for Wide-Field Telescopes}
\label{sec:fo_application}

We have demonstrated that $FO$ is a promising design for various types of X-ray telescopes. This has been shown through the evaluation of four optical designs, each defined by a different set of $\alpha$ and $l$, corresponding to the cases (a) to (d) in Figure \ref{fig:telescopes}.
For all these four optical designs, which cover various types of telescopes, such as soft (Chandra and XMM-Newton) and hard (NuSTAR) X-ray telescopes, the Solar X-ray telescope (Hinode-XRT), and the Survey telescope (eROSITA), it has been demonstrated that the $FO$ design is superior (as shown in Figure \ref{fig:sigma_bar_theta_max} and Table \ref{tab:performance_comparison}). In this section, we specifically discuss the utilization of the $FO$ design for wide-field observations. We briefly discuss three potential science cases where wide-field telescopes with $FO$ designs can be utilized: full-disk imaging spectroscopy of the Sun, wide-field X-ray surveys, and follow-up observations of high-energy transients. For the solar X-ray observations, we also present a telescope design utilizing the $FO$ design concept.

\subsection{Science Cases}
\subsubsection{Full-disk Imaging Spectroscopy of the Sun}
In solar physics, one of the most prominent theories to explain the million Kelvin temperature of the solar corona is the hypothesized unobserved nanoflares ubiquitously present in the solar atmosphere \citep{2006SoPh..234...41K, 1988ApJ...330..474P}. To infer their properties, it is necessary to observe the quiescent solar corona and look for a large number of tiny flare events. Such studies are best done with full-disk imaging spectroscopy in X-rays, which increases the probability of observing them, allowing not only detection but also the estimation of their energetics. Full-disk imaging spectroscopic observations also provide avenues for other investigations, such as the onset of the First Ionization Potential bias in hot active region loops \citep{2023ApJ...955..146M}. Telescopes with good resolution over the entire wide FOV would provide more or less uniform sensitivity across the solar disk, making them optimal for these cases. As shown earlier, for such applications, the $FO$ optical design concept is better suited compared to the $W1$ or $WS$ optics.

\subsubsection{Wide-field X-ray Surveys}
Understanding the formation and rapid growth of SMBHs is a key unresolved issue in probing the early phase of the universe. Studying deep AGNs, possibly in their last stage of accreting SMBH seeds present in early-stage galaxies at high redshifts (z $>$ 6), requires highly sensitive telescopes with a flux limit of $\approx 10^{-19} \mathrm{erg/cm²/s}$ in the 0.2-2 keV energy range \citep{2020A&A...642A.184M, 2019JATIS...5b1001G}. Detecting a large number of faint AGNs requires an all-sky survey with very long exposures that reach millions of seconds. These types of survey telescopes also enable the study of how early galaxy groups evolve into today's massive clusters, thereby enhancing our understanding of large-scale structure formation in the universe. Such studies require survey telescopes with large effective area (0.1–1 $\mathrm{m^2}$) and uniform angular resolution of sub-arcsec across a wide FOV (1 $\mathrm{deg^2}$). These telescopes would greatly benefit from an FO-type optical design.

\subsubsection{Follow-up Observations of EMGW Sources and High-Energy Transients}

Short Gamma-Ray Bursts (sGRBs) were thought to be the result of the merger of two neutron stars. This was confirmed by the first simultaneous observation of the gravitational wave event GW170817 and its electromagnetic counterpart, GRB170817A \citep{2017PhRvL.119p1101A}. An optical counterpart of this event was identified a few hours later, enabling detailed follow-up observations of this event across the entire electromagnetic spectrum. However, this remains the only confirmed association. In order to understand the detailed physics of such events, it is essential to carry out follow-up observations of a large number of sGRBs. Quick afterglow observations of long GRBs, thought to result from the core collapse of massive stars, are also very important for understanding the processes and progenitors. The present generation of all-sky GRB monitors, such as Fermi, or proposed missions like Daksha \citep{bhalerao2024science}, can typically localize GRBs within a few degrees. Quick imaging X-ray observations within the error ellipse of such GRBs provide the best opportunity to identify X-ray and optical counterparts of sGRBs and enable detailed follow-up observations. Such observations would require an X-ray imaging telescope with high and uniform angular resolution across a wide FOV. However, since GRB afterglows are likely to be much brighter, the effective area requirement for such telescopes will be relatively moderate compared to deep-sky survey telescopes, which can be accommodated on a more agile satellite. The FO design is ideally suited for such small X-ray telescopes that can quickly identify the X-ray afterglow of these high-energy transient events.

\subsection{Full Disk Imaging Solar X-ray Telescope} \label{sec:solar_telescope}

To illustrate the $FO$ design concept in greater detail for wide-field observations, specifically for the science case of full-disk imaging spectroscopy of the Sun discussed above, we present an optical design for a solar X-ray telescope that utilizes $FO$. The instrument is designed to perform imaging spectroscopy of the quiet Sun over an energy range of 2--8 keV, within a FOV of $\pm 21$ arcminutes.  As the objective is to observe the quiet Sun in soft X-rays, we consider the effective area (excluding detector efficiency) requirement to be 2--4 \(\mathrm{cm^2}\). First, we obtain the optimum geometrical parameters of the telescope to meet the required effective area, and then we evaluate the imaging performance for that telescope design, considering the $WS$ and $FO$ prescriptions for the mirror.

\subsubsection{Effective Area}

To design the telescope to meet the required effective area, we first fixed some parameters based on various constraints. Considering accommodation on the spacecraft bus of the Indian Space Research Organisation (ISRO), the focal length is fixed at 3 m ($x_0$ = 3 m), and for simplicity, the coating material is considered to be a single layer of platinum. In addition to $x_0$, the effective area of the telescope depends on the number of shells and the values of $\alpha$ and $l$ for each shell. Additionally, the effective area ($A_{\text{eff}}$) of the telescope decreases with an increase in energy or field angle.

\begin{figure}[htbp]
    \centering
    \begin{subfigure}{0.45\textwidth}
        \centering
        \includegraphics[width=\textwidth]{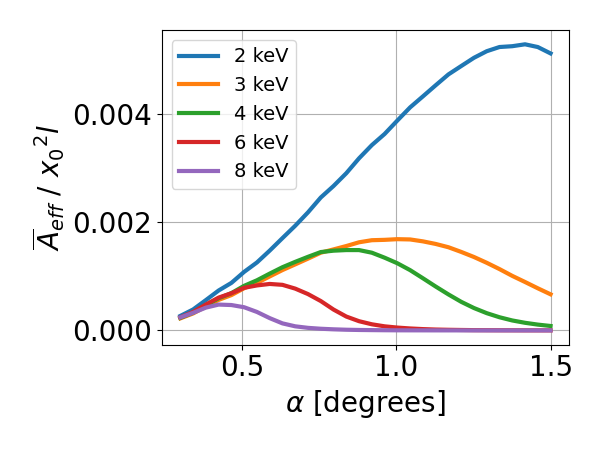}
        \caption{}
        \label{fig:Aeff_alpha}
    \end{subfigure}
    \begin{subfigure}{0.45\textwidth}
        \centering
        \includegraphics[width=\textwidth]{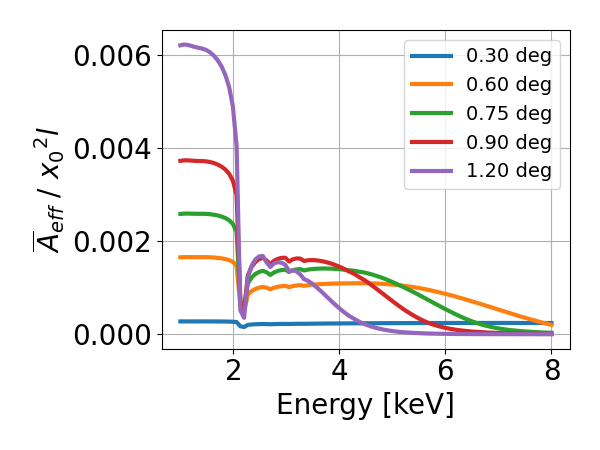}
        \caption{}
        \label{fig:Aeff_Energy}
    \end{subfigure}
    \caption{(a) Averaged and nondimensionalized effective area variation as a function of $\alpha$ for various energy values. (b) Averaged and nondimensionalized effective area variation as a function of energy for various values of $\alpha$.}
    \label{fig:Aeff_xl}
\end{figure}

We define $\overline{A}_{\text{eff}} = \int_{0}^{\theta_{\text{max}}} 2\pi \theta A_{\text{eff}} \, d\theta/\pi \theta_{\text{max}}^2$ as the area-weighted average effective area over the entire FOV. The dependency of $\overline{A}_{\text{eff}}$ on $\alpha$ is not monotonous. The effective area of the telescope is the product of the geometric area and the square of the mirror's reflectivity for incident X-rays. The geometric area increases with $\alpha$, whereas the reflectivity of X-rays decreases with an increase in $\alpha$. Figure \ref{fig:Aeff_alpha} shows the variation of $\overline{A}_{\text{eff}}$ for a single shell, expressed in dimensionless form by dividing $\overline{A}_{\text{eff}}$ by ${x_0}^2$ and $l$, as a function of $\alpha$. It can be seen that the optimal values of $\alpha$ at which $\overline{A}_{\text{eff}}$ is maximized decrease with an increase in X-ray energy. In Figure \ref{fig:Aeff_Energy}, the variation of nondimensionalized $\overline{A}_{\text{eff}}$ as a function of energy for various values of $\alpha$ is shown. Since we are interested in a uniform effective area over the energy range of 2--8 keV, the $\alpha$ that we choose for our telescope design is around $0.6^\circ$.

\begin{figure}[htbp]
\begin{center}
\includegraphics[width=0.55\linewidth]{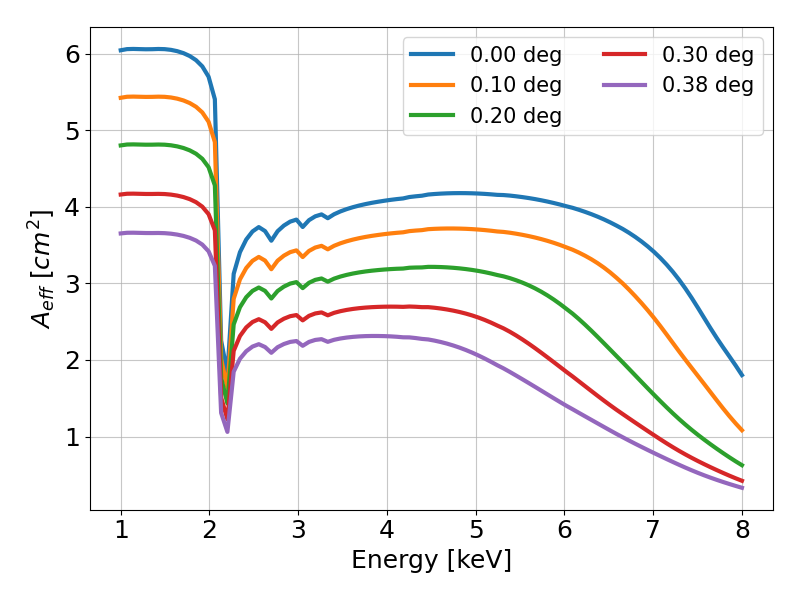}
  \caption{Effective area variation of the three-shell optics with energy for various field angle values, $\theta$.
}
  \label{fig:Eff_Area_Tele}
\end{center}
\end{figure}

Although the effective area can be further improved by increasing $l$, this would decrease the angular resolution (see Equation \ref{Eq:w_empirical} and \ref{Eq:ws_empirical}). Therefore, we restrict $l = 0.01$. To further increase the telescope's effective area to meet the requirements, we included three shells. The parameters of these shells are presented in Table \ref{tab:Tele_parameters} and the variation of $A_{\text{eff}}$ with field angle for different energy values are shown in Figure \ref{fig:Eff_Area_Tele}.

\begin{table}
    \centering
    \begin{tabular}{cccp{2cm}p{2cm}p{1.2cm}}
    \hline
       \textbf{Shell No.}  & \textbf{$\alpha$ (deg)} & \textbf{Radius(mm)} & \textbf{Mirror length (mm)} & \textbf{Shell Thickness (mm)} & \textbf{$x_0$ (mm)} \\
         \hline
        1 & 0.55 & 115.25 & 30 & 5 & 3000 \\
        2 & 0.60 & 125.74 & 30 & 5 & 3000 \\
        3 & 0.65 & 136.23 & 30 & 5 & 3000 \\
        \hline
    \end{tabular}
    \caption{Solar telescope parameters.}
    \label{tab:Tele_parameters}
\end{table}

\subsubsection{Angular Resolution and PSF}

\begin{figure}[!h]
\begin{center}
\includegraphics[width=0.6\linewidth]{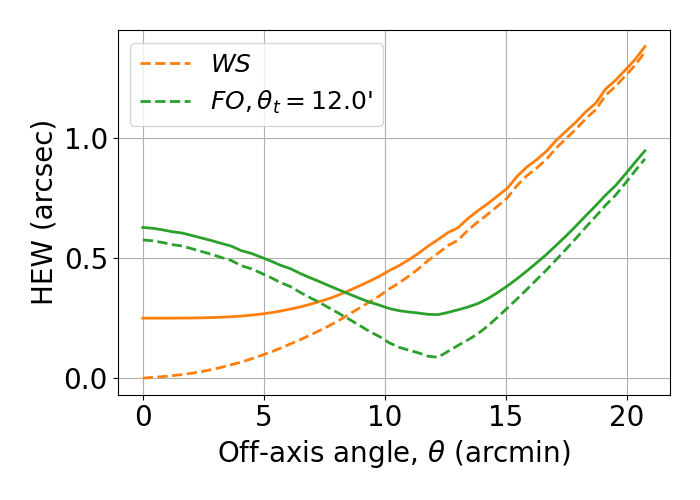}
  \caption{The HEW variation with field angle for a three-shell solar telescope at the curved focal surface is presented for \( WS \) and $FO$ at an energy of 5 keV. The dashed curve represents the HEW obtained without the effect of figure error, while the solid curve accounts for the effect of figure error.}
  \label{fig:AR_Tele}
\end{center}
\end{figure}

To evaluate the imaging performance of the three-shell solar X-ray telescope using $FO$ and $WS$ prescriptions, the half-energy width (HEW, the radius of the PSF containing 50\% of the energy) at a curved focal surface is shown in Figure \ref{fig:AR_Tele}. Figure \ref{fig:AR_Tele} also illustrates the HEW variation while accounting for the artificial figure error present in the $FO$ and \( WS \) prescriptions. The shape of the PSF generated due to figure error is considered Gaussian, with a uniform HEW of 0.25 arcsec for the entire field of view. The \( \overline{\sigma} \) values at the flat and curved focal surfaces, without accounting for figure error, are 0.99 and 0.52 arcsec for \( FO \), and 1.76 and 0.89 arcsec for \( WS \), respectively. $\Delta  {\overline{\sigma}}^2$/$ {\overline{\sigma}}^2$  for $WS$ with respect to $FO$ at the flat and curved focal surfaces are 216\% and 193\%, respectively. Although the operational FOV of the telescope is considered to be 21 arcmin, the \( \theta_{\text{max}} \) for which the average angular resolution for $FO$ is maximized is 17.14 arcmin, which represents the effective size of the solar disc during the quiet phase, containing the majority of the solar flare region. Since the average angular resolution is maximized when \( \theta_{t} \approx 0.7\theta_{\text{max}} \) (see Figure \ref{fig:sigma_bar_theta_max}), \( \theta_t = 12 \) arcmin is considered for all three shells for $FO$.

\begin{figure}
\begin{center}
\includegraphics[clip, trim=0cm 1cm 0cm 0cm, width=0.85\linewidth]{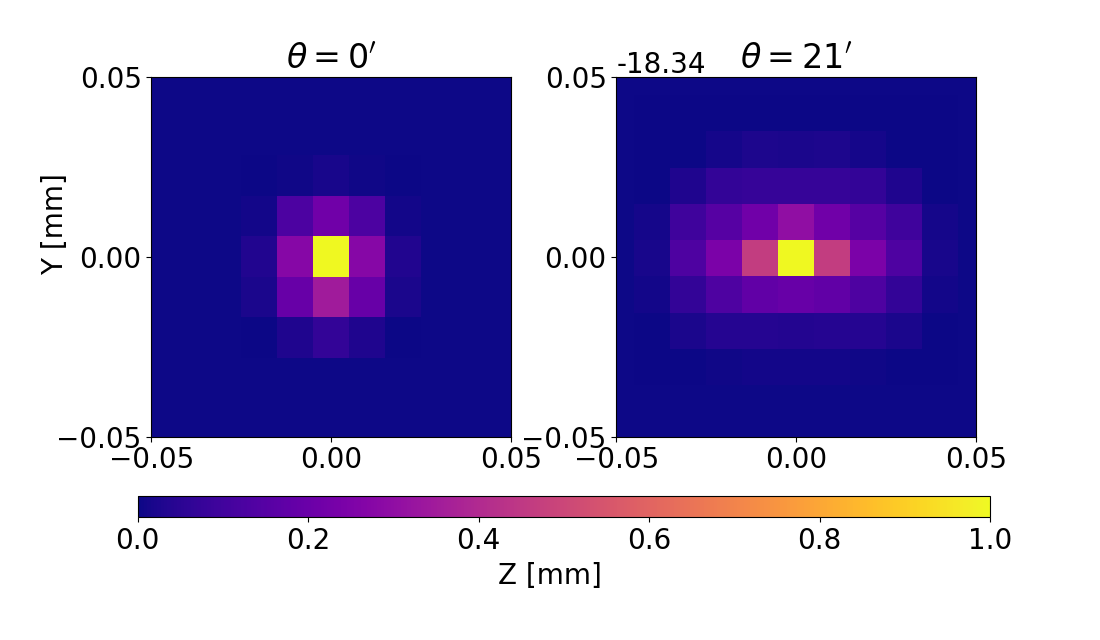}
  \caption{PSF produced by a three-shell solar telescope utilizing $FO$ prescription, at curved focal surfaces for two field angles, $\theta=$ 0 and 21 arcmin, with an energy of 5 keV. The detector pixel size is considered to be $10 \, \mathrm{\mu m}$.}
  \label{fig:psf_Tele}
\end{center}
\end{figure}

The PSF produced by this three-shell telescope with the $FO$ prescription at a curved focal surface is shown in Figure \ref{fig:psf_Tele}. The PSF is estimated for two field angles: $\theta=$ 0 and 21 arcmin. The detector pixel size is 10 µm, and the energy is 5 keV. At a field angle of \( \theta_t = 12 \) arcmin, where the PSF size is minimized, the PSF fits within one pixel.

\begin{figure}[!h]
\begin{center}
\includegraphics[clip, trim=0cm 0cm 0cm 0cm, width=1\linewidth]{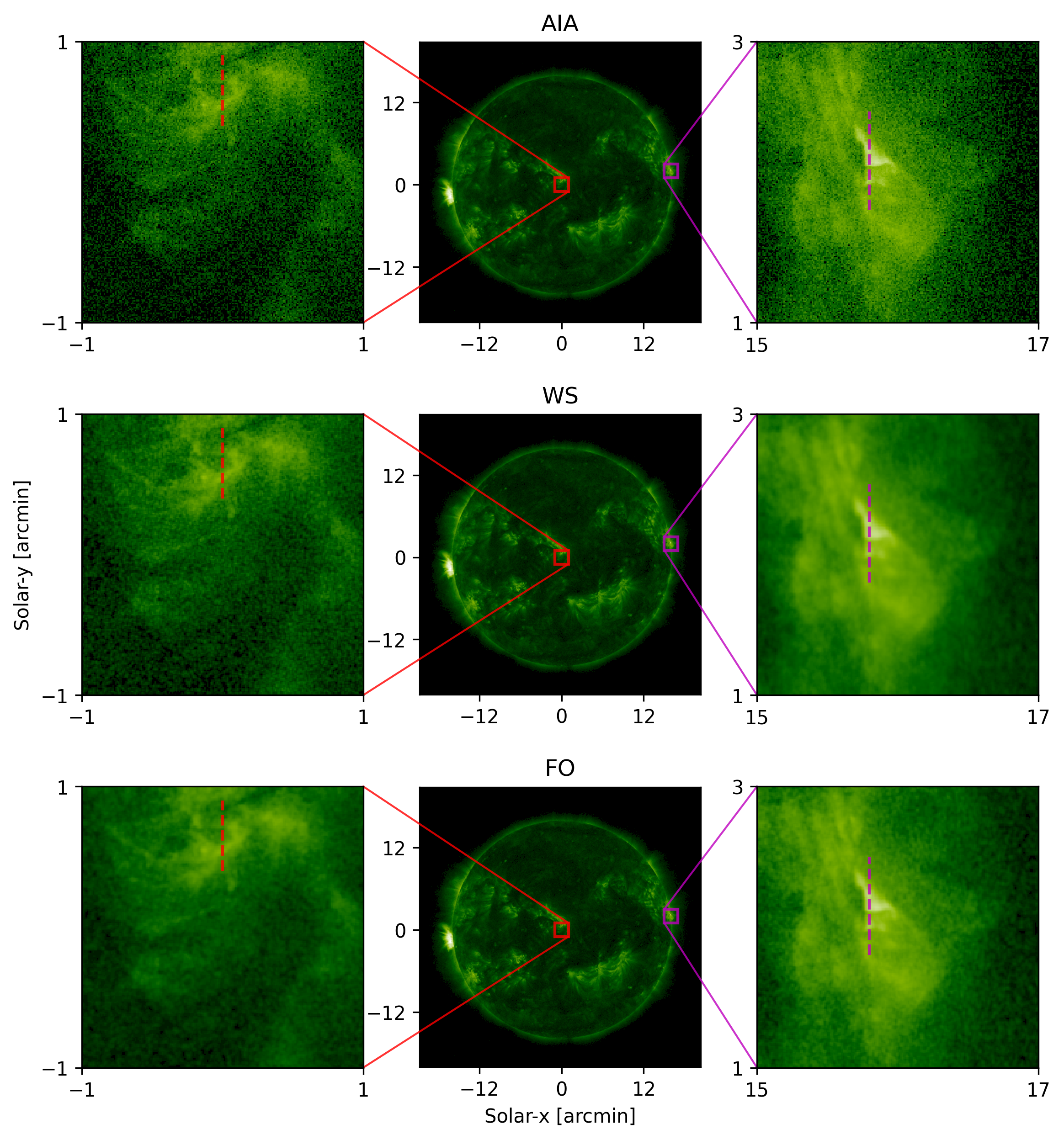}
  \caption{The first row represents the image of the Sun obtained from SDO/AIA at 94 \r{A}, after being deconvolved with the instrument's PSF. The second and third rows show the AIA images convolved with the PSFs of $WS$ and $FO$, respectively, obtained at the curved focal surface. Each image includes two zoomed regions: the central region of the Sun on the left side and the solar limb region on the right side.}
  \label{fig:SDO_AIA_SUN}
\end{center}
\end{figure}

\begin{figure}[!h]
\begin{center}
\includegraphics[clip, trim=0cm 0cm 0cm 0cm, width=0.85\linewidth]{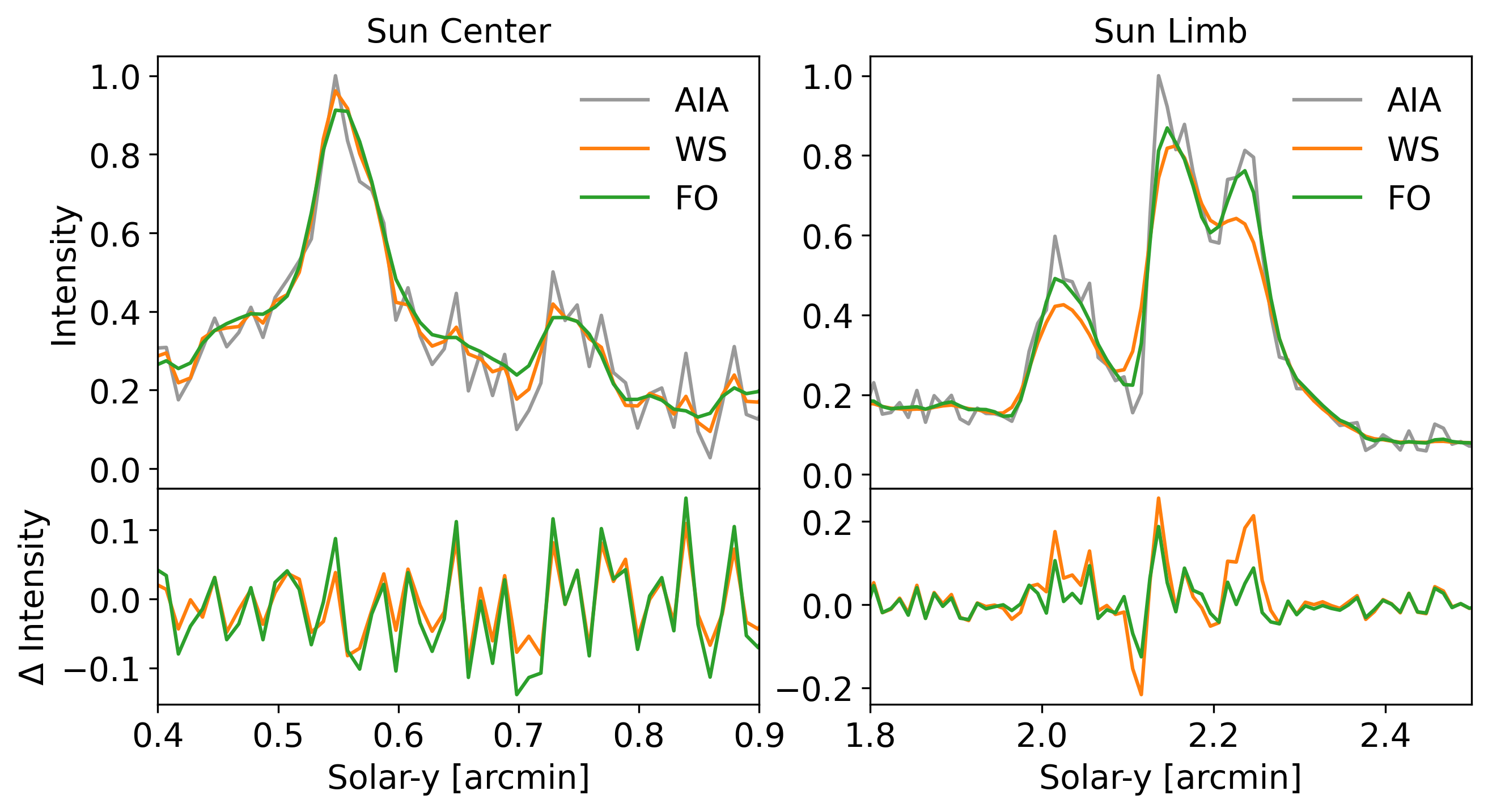}
  \caption{Intensity variation along a line in the central and limb regions of the Sun. The lines, marked in red and magenta in the zoomed-in regions of Figure \ref{fig:SDO_AIA_SUN}, indicate where the intensity variation is measured. The lower panel shows the intensity differences between AIA and WS, and AIA and FO.}
  \label{fig:SDO_AIA_SUN_1D}
\end{center}
\end{figure}

At average scale, $FO$ achieves better angular resolution than $WS$ because it minimizes the PSF size at far-field angles, covering a larger area of the field of view, whereas $WS$ minimizes PSF size only at the center, limiting its effectiveness to a small region. To illustrate this comparison, Figure \ref{fig:SDO_AIA_SUN} shows the convolved image of the Sun with the three-shell solar telescope, where the telescope's PSF is generated using $WS$ and $FO$ prescriptions at the curved focal surface. The raw image of the Sun was obtained from the Solar Dynamics Observatory/Atmospheric Imaging Assembly (SDO/AIA; \cite{lemen2012atmospheric}) instrument, captured in the 94 Å filter on August 27, 2023. It has been deconvolved with the AIA's instrument PSF at 94 Å. The solar images from the AIA in the 94 \AA\ filter are chosen because this instrument is developed by utilizing normal incident optics for the extreme ultraviolet energy band. Thus, it provides high-resolution images of the entire solar disk, and the 94 \AA\ filter is the closest to the soft X-ray wavelength. The AIA image covers a $41' \times 41'$ FOV sampled with $\mathrm{4k} \times \mathrm{4k}$ pixels. Since the shape and size of the PSF of $FO$ and $WS$ vary within the FOV, the $41' \times 41'$ FOV is divided into $\mathrm{128} \times \mathrm{128}$ square images. Each square image is convolved with a PSF corresponding to the center of the square and stitched together to construct the whole image. The orientation of the PSF is also accounted for in each square image based on the azimuthal location ($\phi$) of its center, in addition to the radial position ($\theta$). Additionally, to account for the effect of figure error, the PSF of $FO$ and $WS$ is convolved with a Gaussian PSF having a HEW of 0.25 arcsec. The image quality of the zoomed central region of the Sun ($2' \times 2'$) is degraded for both $WS$ and $FO$ when compared to the AIA image. However, the degradation in the central region is higher in $FO$ compared to $WS$. The zoomed images of the region at the far field angle (limb region) show that both $WS$ and $FO$ degrade compared to the AIA image, but the degradation in $WS$ is higher than in $FO$. However, degradation in the central region for $FO$ has a lesser impact compared to degradation at far-field angles for $WS$ when evaluating imaging performance on an average scale.

The superior performance of $FO$ in the limb region and relatively lower performance in the central region compared to $WS$ is once again shown in Figure \ref{fig:SDO_AIA_SUN_1D} through the intensity variation along the line. The lines along which the intensity variation is shown are marked in the zoomed-in region of Figure \ref{fig:SDO_AIA_SUN}, with red for the central region and magenta for the limb region. We selected regions with high-intensity flares. It can be seen that in the central region, the $WS$ curve matches more closely with AIA, whereas in the limb region, the $FO$ curve matches more closely with AIA. In the limb region, $WS$ shows a smoother trend compared to $FO$, causing $WS$ to lose more information about the sharp peaks present in AIA. When comparing the differences between AIA and $WS$, and AIA and $FO$ (shown in the lower panel), deviations are higher in $WS$ than in $FO$, particularly in the limb region, especially in the flaring area.

\section{Summary}
We introduced a new optical design named Field-Angle Optimized ($FO$) to optimize the angular resolution of wide-field X-ray telescopes. $FO$ demonstrates improved imaging performance compared to other existing optical designs, such as $W1$, $WS$, and $HH$, when considered over an averaged scale for the entire field of view. $FO$ provides coma-free imaging, much like $WS$, and while the angular resolution in $WS$ is maximized along the axial direction, $FO$ offers the option to maximize the angular resolution at any field angle. When $FO$ is optimized for the axial direction, it yields the same result as $WS$. On average, the improvement of $FO$ compared to $W1$ for various practical telescope designs ranges from 85\% to 208\% at a flat focal surface and from 217\% to 648\% at a curved focal surface. The improvement with respect to $WS$ is around 64\% to 97\% at a flat focal surface and around 167\% to 294\% at a curved focal surface. From a fabrication perspective, the surface figure of $FO$ is very similar to $W1$ and $WS$ and is feasible from a realization standpoint. Finally, to demonstrate the use of the $FO$ design, we utilized it to provide a design for a solar X-ray telescope for imaging and spectroscopy. This telescope design is optimized as a three-shell telescope for an energy range of 2-8 keV and a field of view of $\pm$21 arcmin. The angular resolution for this telescope is maximized at a field angle of 12 arcmin by utilizing $FO$. The average rms image radius of this telescope is 0.99 arcsec at a flat focal surface and 0.52 arcsec at a curved focal surface. It has been shown that the $FO$ design provides a significant advantage over existing optical designs, not only in the context of solar X-ray telescopes but also across a broader range of X-ray telescope applications.

\bmhead{Acknowledgements} This work at the Physical Research Laboratory, Ahmedabad, is supported by the Department of Space, Government of India.

\begin{appendices}
\section{$FO$ Prescription Calculation}\label{secA1}

To begin, we first require some geometrical parameters of the mirror, such as $\alpha$, $l$, $x_0$, and $\theta_t$, for which the angular resolution needs to be maximized. Using $x_0$ and $\alpha$, the intersection point, $(x_0, y_0)$, of the two mirrors at the IP can be determined. Starting from the IP (see Figure \ref{fig:prescription}), both mirrors can be extended to the desired length using the following steps.

Step 1: Building the first segment of the secondary mirror, $g_1(x_2)$.

Consider a starting point on the primary mirror, very close to the IP, with a slope angle of $\alpha$. Next, estimate a starting point on the secondary mirror, also near the IP, with a slope angle of $3\alpha$, using the constraint that a ray at a field angle of $\theta_t$, passing through these two points, reaches the focal plane at $y = -f\theta_t$.

From the starting point on the secondary mirror, move a very small distance (infinitesimal step) toward the negative x-direction along the slope of $3\alpha$. The next task is to find the new slope at this point on the secondary mirror, which can be done in two steps. First, determine the field angle $\theta$ at which a ray striking the primary mirror's starting point reaches the secondary mirror at this new position. Once $\theta$ is found, the slope at this point can be estimated using the constraint that the ray hitting the primary mirror with a field angle of $\theta$, after reflecting from the secondary mirror, reaches the focal plane at $y = -f\theta$.

Next, move again on the secondary mirror with a very small step size along the new slope. At this new point, the task is to find the slope angle again, applying the same constraint as before: rays from the starting point on the primary mirror will reach the secondary mirror at this point for a specific field angle $\theta$, and the same ray will then reach the focal plane at $y = -f\theta$. This process is repeated iteratively, building the secondary mirror step by step, until it reaches a length where the specific field angle $\theta$, which started as $\theta_t$, reaches $-\theta_t$.

Step 2: Obtaining the first segment of the primary mirror, $f_1(x_1)$, by utilizing $g_1(x_2)$.

From the starting point on the primary mirror, move a very small step in the positive x-direction along the slope angle $\alpha$. The updated slope at this new position can be determined by applying the constraint that a ray, at a field angle of $\theta_t$ from this point, follows the reflection from the first segment of the secondary mirror and reaches the focal plane at $y = -f\theta_t$. After obtaining the updated slope, move a small step again in the positive x-direction along the new slope, and revise the slope at the new position using the same constraint.
By repeating this process, the prescription of the first segment of the primary mirror can be determined up to a length where a ray at $\theta_t$ from its edge (towards +x) meets the edge (towards -x) of the first segment of the secondary mirror.

Step 3: Obtaining the second segment of the secondary mirror, $g_2(x_2)$, by utilizing $f_1(x_1)$.

From the edge (towards -x) of the first segment of the secondary mirror, move a very small step towards -x along its slope. The updated slope at this new point can be determined by applying the constraint that a ray with a field angle of $-\theta_t$, passing through this point after being reflected from the first segment of the primary mirror, reaches the focal plane at $y = f\theta_t$. By continuing to move towards -x and applying the same constraint, the slope can be updated further, and the curve can be extended until the edge (towards -x) of this segment receives the ray from the edge (towards +x) of the first segment of the primary mirror at $-\theta_t$.

Similarly, $f_2(x_1)$ and $g_3(x_2)$ can be determined using $g_2(x_2)$, and so on. Both mirrors can be built in this way until the length of both reaches $x_0l$.

\end{appendices}

\bibliography{references}


\begin{thebibliography}{53}
\ifx \bisbn   \undefined \def \bisbn  #1{ISBN #1}\fi
\ifx \binits  \undefined \def \binits#1{#1}\fi
\ifx \bauthor  \undefined \def \bauthor#1{#1}\fi
\ifx \batitle  \undefined \def \batitle#1{#1}\fi
\ifx \bjtitle  \undefined \def \bjtitle#1{#1}\fi
\ifx \bvolume  \undefined \def \bvolume#1{\textbf{#1}}\fi
\ifx \byear  \undefined \def \byear#1{#1}\fi
\ifx \bissue  \undefined \def \bissue#1{#1}\fi
\ifx \bfpage  \undefined \def \bfpage#1{#1}\fi
\ifx \blpage  \undefined \def \blpage #1{#1}\fi
\ifx \burl  \undefined \def \burl#1{\textsf{#1}}\fi
\ifx \doiurl  \undefined \def \doiurl#1{\url{https://doi.org/#1}}\fi
\ifx \betal  \undefined \def \betal{\textit{et al.}}\fi
\ifx \binstitute  \undefined \def \binstitute#1{#1}\fi
\ifx \binstitutionaled  \undefined \def \binstitutionaled#1{#1}\fi
\ifx \bctitle  \undefined \def \bctitle#1{#1}\fi
\ifx \beditor  \undefined \def \beditor#1{#1}\fi
\ifx \bpublisher  \undefined \def \bpublisher#1{#1}\fi
\ifx \bbtitle  \undefined \def \bbtitle#1{#1}\fi
\ifx \bedition  \undefined \def \bedition#1{#1}\fi
\ifx \bseriesno  \undefined \def \bseriesno#1{#1}\fi
\ifx \blocation  \undefined \def \blocation#1{#1}\fi
\ifx \bsertitle  \undefined \def \bsertitle#1{#1}\fi
\ifx \bsnm \undefined \def \bsnm#1{#1}\fi
\ifx \bsuffix \undefined \def \bsuffix#1{#1}\fi
\ifx \bparticle \undefined \def \bparticle#1{#1}\fi
\ifx \barticle \undefined \def \barticle#1{#1}\fi
\bibcommenthead
\ifx \bconfdate \undefined \def \bconfdate #1{#1}\fi
\ifx \botherref \undefined \def \botherref #1{#1}\fi
\ifx \url \undefined \def \url#1{\textsf{#1}}\fi
\ifx \bchapter \undefined \def \bchapter#1{#1}\fi
\ifx \bbook \undefined \def \bbook#1{#1}\fi
\ifx \bcomment \undefined \def \bcomment#1{#1}\fi
\ifx \oauthor \undefined \def \oauthor#1{#1}\fi
\ifx \citeauthoryear \undefined \def \citeauthoryear#1{#1}\fi
\ifx \endbibitem  \undefined \def \endbibitem {}\fi
\ifx \bconflocation  \undefined \def \bconflocation#1{#1}\fi
\ifx \arxivurl  \undefined \def \arxivurl#1{\textsf{#1}}\fi
\csname PreBibitemsHook\endcsname

\bibitem[\protect\citeauthoryear{{Abbott} et~al.}{2017}]{2017PhRvL.119p1101A}
\begin{barticle}
\bauthor{\bsnm{{Abbott}}, \binits{B.P.}},
\bauthor{\bsnm{{Abbott}}, \binits{R.}},
\bauthor{\bsnm{{Abbott}}, \binits{T.D.}},
\bauthor{\bsnm{{Acernese}}, \binits{F.}},
\bauthor{\bsnm{{Ackley}}, \binits{K.}},
\bauthor{\bsnm{{Adams}}, \binits{C.}},
\bauthor{\bsnm{{Adams}}, \binits{T.}},
\bauthor{\bsnm{{Addesso}}, \binits{P.}},
\bauthor{\bsnm{{Adhikari}}, \binits{R.X.}},
\bauthor{\bsnm{{Adya}}, \binits{V.B.}},
\bauthor{\bsnm{{Affeldt}}, \binits{C.}},
\bauthor{\bsnm{{Afrough}}, \binits{M.}},
\bauthor{\bsnm{{Agarwal}}, \binits{B.}},
\bauthor{\bsnm{{Agathos}}, \binits{M.}},
\bauthor{\bsnm{{Agatsuma}}, \binits{K.}},
\bauthor{\bsnm{{Aggarwal}}, \binits{N.}},
\bauthor{\bsnm{{Aguiar}}, \binits{O.D.}},
\bauthor{\bsnm{{Aiello}}, \binits{L.}},
\bauthor{\bsnm{{Ain}}, \binits{A.}},
\bauthor{\bsnm{{Ajith}}, \binits{P.}},
\bauthor{\bsnm{{Allen}}, \binits{B.}},
\bauthor{\bsnm{{Allen}}, \binits{G.}},
\bauthor{\bsnm{{Allocca}}, \binits{A.}},
\bauthor{\bsnm{{Altin}}, \binits{P.A.}},
\bauthor{\bsnm{{Amato}}, \binits{A.}},
\bauthor{\bsnm{{Ananyeva}}, \binits{A.}},
\bauthor{\bsnm{{Anderson}}, \binits{S.B.}},
\bauthor{\bsnm{{Anderson}}, \binits{W.G.}},
\bauthor{\bsnm{{Angelova}}, \binits{S.V.}},
\bauthor{\bsnm{{Antier}}, \binits{S.}},
\bauthor{\bsnm{{Appert}}, \binits{S.}},
\bauthor{\bsnm{{Arai}}, \binits{K.}},
\bauthor{\bsnm{{Araya}}, \binits{M.C.}},
\bauthor{\bsnm{{Areeda}}, \binits{J.S.}},
\bauthor{\bsnm{{Arnaud}}, \binits{N.}},
\bauthor{\bsnm{{Arun}}, \binits{K.G.}},
\bauthor{\bsnm{{Ascenzi}}, \binits{S.}},
\bauthor{\bsnm{{Ashton}}, \binits{G.}},
\bauthor{\bsnm{{Ast}}, \binits{M.}},
\bauthor{\bsnm{{Aston}}, \binits{S.M.}},
\bauthor{\bsnm{{Astone}}, \binits{P.}},
\bauthor{\bsnm{{Atallah}}, \binits{D.V.}},
\bauthor{\bsnm{{Aufmuth}}, \binits{P.}},
\bauthor{\bsnm{{Aulbert}}, \binits{C.}},
\bauthor{\bsnm{{AultONeal}}, \binits{K.}},
\bauthor{\bsnm{{Austin}}, \binits{C.}},
\bauthor{\bsnm{{Avila-Alvarez}}, \binits{A.}},
\bauthor{\bsnm{{Babak}}, \binits{S.}},
\bauthor{\bsnm{{Bacon}}, \binits{P.}},
\bauthor{\bsnm{{Bader}}, \binits{M.K.M.}},
\bauthor{\bsnm{{Bae}}, \binits{S.}},
\bauthor{\bsnm{{Bailes}}, \binits{M.}},
\bauthor{\bsnm{{Baker}}, \binits{P.T.}},
\bauthor{\bsnm{{Baldaccini}}, \binits{F.}},
\bauthor{\bsnm{{Ballardin}}, \binits{G.}},
\bauthor{\bsnm{{Ballmer}}, \binits{S.W.}},
\bauthor{\bsnm{{Banagiri}}, \binits{S.}},
\bauthor{\bsnm{{Barayoga}}, \binits{J.C.}},
\bauthor{\bsnm{{Barclay}}, \binits{S.E.}},
\bauthor{\bsnm{{Barish}}, \binits{B.C.}},
\bauthor{\bsnm{{Barker}}, \binits{D.}},
\bauthor{\bsnm{{Barkett}}, \binits{K.}},
\bauthor{\bsnm{{Barone}}, \binits{F.}},
\bauthor{\bsnm{{Barr}}, \binits{B.}},
\bauthor{\bsnm{{Barsotti}}, \binits{L.}},
\bauthor{\bsnm{{Barsuglia}}, \binits{M.}},
\bauthor{\bsnm{{Barta}}, \binits{D.}},
\bauthor{\bsnm{{Barthelmy}}, \binits{S.D.}},
\bauthor{\bsnm{{Bartlett}}, \binits{J.}},
\bauthor{\bsnm{{Bartos}}, \binits{I.}},
\bauthor{\bsnm{{Bassiri}}, \binits{R.}},
\bauthor{\bsnm{{Basti}}, \binits{A.}},
\bauthor{\bsnm{{Batch}}, \binits{J.C.}},
\bauthor{\bsnm{{Bawaj}}, \binits{M.}},
\bauthor{\bsnm{{Bayley}}, \binits{J.C.}},
\bauthor{\bsnm{{Bazzan}}, \binits{M.}},
\bauthor{\bsnm{{B{\'e}csy}}, \binits{B.}},
\bauthor{\bsnm{{Beer}}, \binits{C.}},
\bauthor{\bsnm{{Bejger}}, \binits{M.}},
\bauthor{\bsnm{{Belahcene}}, \binits{I.}},
\bauthor{\bsnm{{Bell}}, \binits{A.S.}},
\bauthor{\bsnm{{Berger}}, \binits{B.K.}},
\bauthor{\bsnm{{Bergmann}}, \binits{G.}},
\bauthor{\bsnm{{Bernuzzi}}, \binits{S.}},
\bauthor{\bsnm{{Bero}}, \binits{J.J.}},
\bauthor{\bsnm{{Berry}}, \binits{C.P.L.}},
\bauthor{\bsnm{{Bersanetti}}, \binits{D.}},
\bauthor{\bsnm{{Bertolini}}, \binits{A.}},
\bauthor{\bsnm{{Betzwieser}}, \binits{J.}},
\bauthor{\bsnm{{Bhagwat}}, \binits{S.}},
\bauthor{\bsnm{{Bhandare}}, \binits{R.}},
\bauthor{\bsnm{{Bilenko}}, \binits{I.A.}},
\bauthor{\bsnm{{Billingsley}}, \binits{G.}},
\bauthor{\bsnm{{Billman}}, \binits{C.R.}},
\bauthor{\bsnm{{Birch}}, \binits{J.}},
\bauthor{\bsnm{{Birney}}, \binits{R.}},
\bauthor{\bsnm{{Birnholtz}}, \binits{O.}},
\bauthor{\bsnm{{Biscans}}, \binits{S.}},
\bauthor{\bsnm{{Biscoveanu}}, \binits{S.}},
\bauthor{\bsnm{{Bisht}}, \binits{A.}},
\bauthor{\bsnm{{Bitossi}}, \binits{M.}},
\bauthor{\bsnm{{Biwer}}, \binits{C.}},
\bauthor{\bsnm{{Bizouard}}, \binits{M.A.}},
\bauthor{\bsnm{{Blackburn}}, \binits{J.K.}},
\bauthor{\bsnm{{Blackman}}, \binits{J.}},
\bauthor{\bsnm{{Blair}}, \binits{C.D.}},
\bauthor{\bsnm{{Blair}}, \binits{D.G.}},
\bauthor{\bsnm{{Blair}}, \binits{R.M.}},
\bauthor{\bsnm{{Bloemen}}, \binits{S.}},
\bauthor{\bsnm{{Bock}}, \binits{O.}},
\bauthor{\bsnm{{Bode}}, \binits{N.}},
\bauthor{\bsnm{{Boer}}, \binits{M.}},
\bauthor{\bsnm{{Bogaert}}, \binits{G.}},
\bauthor{\bsnm{{Bohe}}, \binits{A.}},
\bauthor{\bsnm{{Bondu}}, \binits{F.}},
\bauthor{\bsnm{{Bonilla}}, \binits{E.}},
\bauthor{\bsnm{{Bonnand}}, \binits{R.}},
\bauthor{\bsnm{{Boom}}, \binits{B.A.}},
\bauthor{\bsnm{{Bork}}, \binits{R.}},
\bauthor{\bsnm{{Boschi}}, \binits{V.}},
\bauthor{\bsnm{{Bose}}, \binits{S.}},
\bauthor{\bsnm{{Bossie}}, \binits{K.}},
\bauthor{\bsnm{{Bouffanais}}, \binits{Y.}},
\bauthor{\bsnm{{Bozzi}}, \binits{A.}},
\bauthor{\bsnm{{Bradaschia}}, \binits{C.}},
\bauthor{\bsnm{{Brady}}, \binits{P.R.}},
\bauthor{\bsnm{{Branchesi}}, \binits{M.}},
\bauthor{\bsnm{{Brau}}, \binits{J.E.}},
\bauthor{\bsnm{{Briant}}, \binits{T.}},
\bauthor{\bsnm{{Brillet}}, \binits{A.}},
\bauthor{\bsnm{{Brinkmann}}, \binits{M.}},
\bauthor{\bsnm{{Brisson}}, \binits{V.}},
\bauthor{\bsnm{{Brockill}}, \binits{P.}},
\bauthor{\bsnm{{Broida}}, \binits{J.E.}},
\bauthor{\bsnm{{Brooks}}, \binits{A.F.}},
\bauthor{\bsnm{{Brown}}, \binits{D.A.}},
\bauthor{\bsnm{{Brown}}, \binits{D.D.}},
\bauthor{\bsnm{{Brunett}}, \binits{S.}},
\bauthor{\bsnm{{Buchanan}}, \binits{C.C.}},
\bauthor{\bsnm{{Buikema}}, \binits{A.}},
\bauthor{\bsnm{{Bulik}}, \binits{T.}},
\bauthor{\bsnm{{Bulten}}, \binits{H.J.}},
\bauthor{\bsnm{{Buonanno}}, \binits{A.}},
\bauthor{\bsnm{{Buskulic}}, \binits{D.}},
\bauthor{\bsnm{{Buy}}, \binits{C.}},
\bauthor{\bsnm{{Byer}}, \binits{R.L.}},
\bauthor{\bsnm{{Cabero}}, \binits{M.}},
\bauthor{\bsnm{{Cadonati}}, \binits{L.}},
\bauthor{\bsnm{{Cagnoli}}, \binits{G.}},
\bauthor{\bsnm{{Cahillane}}, \binits{C.}},
\bauthor{\bsnm{{Calder{\'o}n Bustillo}}, \binits{J.}},
\bauthor{\bsnm{{Callister}}, \binits{T.A.}},
\bauthor{\bsnm{{Calloni}}, \binits{E.}},
\bauthor{\bsnm{{Camp}}, \binits{J.B.}},
\bauthor{\bsnm{{Canepa}}, \binits{M.}},
\bauthor{\bsnm{{Canizares}}, \binits{P.}},
\bauthor{\bsnm{{Cannon}}, \binits{K.C.}},
\bauthor{\bsnm{{Cao}}, \binits{H.}},
\bauthor{\bsnm{{Cao}}, \binits{J.}},
\bauthor{\bsnm{{Capano}}, \binits{C.D.}},
\bauthor{\bsnm{{Capocasa}}, \binits{E.}},
\bauthor{\bsnm{{Carbognani}}, \binits{F.}},
\bauthor{\bsnm{{Caride}}, \binits{S.}},
\bauthor{\bsnm{{Carney}}, \binits{M.F.}},
\bauthor{\bsnm{{Carullo}}, \binits{G.}},
\bauthor{\bsnm{{Casanueva Diaz}}, \binits{J.}},
\bauthor{\bsnm{{Casentini}}, \binits{C.}},
\bauthor{\bsnm{{Caudill}}, \binits{S.}},
\bauthor{\bsnm{{Cavagli{\`a}}}, \binits{M.}},
\bauthor{\bsnm{{Cavalier}}, \binits{F.}},
\bauthor{\bsnm{{Cavalieri}}, \binits{R.}},
\bauthor{\bsnm{{Cella}}, \binits{G.}},
\bauthor{\bsnm{{Cepeda}}, \binits{C.B.}},
\bauthor{\bsnm{{Cerd{\'a}-Dur{\'a}n}}, \binits{P.}},
\bauthor{\bsnm{{Cerretani}}, \binits{G.}},
\bauthor{\bsnm{{Cesarini}}, \binits{E.}},
\bauthor{\bsnm{{Chamberlin}}, \binits{S.J.}},
\bauthor{\bsnm{{Chan}}, \binits{M.}},
\bauthor{\bsnm{{Chao}}, \binits{S.}},
\bauthor{\bsnm{{Charlton}}, \binits{P.}},
\bauthor{\bsnm{{Chase}}, \binits{E.}},
\bauthor{\bsnm{{Chassande-Mottin}}, \binits{E.}},
\bauthor{\bsnm{{Chatterjee}}, \binits{D.}},
\bauthor{\bsnm{{Chatziioannou}}, \binits{K.}},
\bauthor{\bsnm{{Cheeseboro}}, \binits{B.D.}},
\bauthor{\bsnm{{Chen}}, \binits{H.Y.}},
\bauthor{\bsnm{{Chen}}, \binits{X.}},
\bauthor{\bsnm{{Chen}}, \binits{Y.}},
\bauthor{\bsnm{{Cheng}}, \binits{H.-P.}},
\bauthor{\bsnm{{Chia}}, \binits{H.}},
\bauthor{\bsnm{{Chincarini}}, \binits{A.}},
\bauthor{\bsnm{{Chiummo}}, \binits{A.}},
\bauthor{\bsnm{{Chmiel}}, \binits{T.}},
\bauthor{\bsnm{{Cho}}, \binits{H.S.}},
\bauthor{\bsnm{{Cho}}, \binits{M.}},
\bauthor{\bsnm{{Chow}}, \binits{J.H.}},
\bauthor{\bsnm{{Christensen}}, \binits{N.}},
\bauthor{\bsnm{{Chu}}, \binits{Q.}},
\bauthor{\bsnm{{Chua}}, \binits{A.J.K.}},
\bauthor{\bsnm{{Chua}}, \binits{S.}}:
\batitle{{GW170817: Observation of Gravitational Waves from a Binary Neutron
  Star Inspiral}}.
\bjtitle{\prl}
\bvolume{119}(\bissue{16}),
\bfpage{161101}
(\byear{2017})
\doiurl{10.1103/PhysRevLett.119.161101}
{\href{https://arxiv.org/abs/1710.05832}{{arXiv:1710.05832}}}
{[gr-qc]}
\end{barticle}
\endbibitem

\bibitem[\protect\citeauthoryear{Aschenbach}{1985}]{aschenbach1985x}
\begin{barticle}
\bauthor{\bsnm{Aschenbach}, \binits{B.}}:
\batitle{X-ray telescopes}.
\bjtitle{Reports on Progress in Physics}
\bvolume{48}(\bissue{5}),
\bfpage{579}
(\byear{1985})
\end{barticle}
\endbibitem

\bibitem[\protect\citeauthoryear{{Aschenbach}}{1988}]{aschenbach1988design}
\begin{barticle}
\bauthor{\bsnm{{Aschenbach}}, \binits{B.}}:
\batitle{{Design, construction, and performance of the ROSAT high- resolution
  x-ray mirror assembly}}.
\bjtitle{\ao}
\bvolume{27}(\bissue{8}),
\bfpage{1404}--\blpage{1413}
(\byear{1988})
\doiurl{10.1364/AO.27.001404}
\end{barticle}
\endbibitem

\bibitem[\protect\citeauthoryear{{Burrows} et~al.}{1992}]{burrows1992optimal}
\begin{barticle}
\bauthor{\bsnm{{Burrows}}, \binits{C.J.}},
\bauthor{\bsnm{{Burg}}, \binits{R.}},
\bauthor{\bsnm{{Giacconi}}, \binits{R.}}:
\batitle{{Optimal Grazing Incidence Optics and Its Application to Wide-Field
  X-Ray Imaging}}.
\bjtitle{\apj}
\bvolume{392},
\bfpage{760}
(\byear{1992})
\doiurl{10.1086/171479}
\end{barticle}
\endbibitem

\bibitem[\protect\citeauthoryear{{Burrows} et~al.}{2005}]{burrows2005swift}
\begin{barticle}
\bauthor{\bsnm{{Burrows}}, \binits{D.N.}},
\bauthor{\bsnm{{Hill}}, \binits{J.E.}},
\bauthor{\bsnm{{Nousek}}, \binits{J.A.}},
\bauthor{\bsnm{{Kennea}}, \binits{J.A.}},
\bauthor{\bsnm{{Wells}}, \binits{A.}},
\bauthor{\bsnm{{Osborne}}, \binits{J.P.}},
\bauthor{\bsnm{{Abbey}}, \binits{A.F.}},
\bauthor{\bsnm{{Beardmore}}, \binits{A.}},
\bauthor{\bsnm{{Mukerjee}}, \binits{K.}},
\bauthor{\bsnm{{Short}}, \binits{A.D.T.}},
\bauthor{\bsnm{{Chincarini}}, \binits{G.}},
\bauthor{\bsnm{{Campana}}, \binits{S.}},
\bauthor{\bsnm{{Citterio}}, \binits{O.}},
\bauthor{\bsnm{{Moretti}}, \binits{A.}},
\bauthor{\bsnm{{Pagani}}, \binits{C.}},
\bauthor{\bsnm{{Tagliaferri}}, \binits{G.}},
\bauthor{\bsnm{{Giommi}}, \binits{P.}},
\bauthor{\bsnm{{Capalbi}}, \binits{M.}},
\bauthor{\bsnm{{Tamburelli}}, \binits{F.}},
\bauthor{\bsnm{{Angelini}}, \binits{L.}},
\bauthor{\bsnm{{Cusumano}}, \binits{G.}},
\bauthor{\bsnm{{Br{\"a}uninger}}, \binits{H.W.}},
\bauthor{\bsnm{{Burkert}}, \binits{W.}},
\bauthor{\bsnm{{Hartner}}, \binits{G.D.}}:
\batitle{{The Swift X-Ray Telescope}}.
\bjtitle{\ssr}
\bvolume{120}(\bissue{3-4}),
\bfpage{165}--\blpage{195}
(\byear{2005})
\doiurl{10.1007/s11214-005-5097-2}
{\href{https://arxiv.org/abs/astro-ph/0508071}{{arXiv:astro-ph/0508071}}}
{[astro-ph]}
\end{barticle}
\endbibitem

\bibitem[\protect\citeauthoryear{Bhalerao et~al.}{2024a}]{bhalerao2024science}
\begin{barticle}
\bauthor{\bsnm{Bhalerao}, \binits{V.}},
\bauthor{\bsnm{Sawant}, \binits{D.}},
\bauthor{\bsnm{Pai}, \binits{A.}},
\bauthor{\bsnm{Tendulkar}, \binits{S.}},
\bauthor{\bsnm{Vadawale}, \binits{S.}},
\bauthor{\bsnm{Bhattacharya}, \binits{D.}},
\bauthor{\bsnm{Rana}, \binits{V.}},
\bauthor{\bsnm{Adalja}, \binits{H.K.L.}},
\bauthor{\bsnm{Anupama}, \binits{G.}},
\bauthor{\bsnm{Bala}, \binits{S.}}, \betal:
\batitle{{Science with the Daksha high energy transients mission}}.
\bjtitle{Experimental Astronomy}
\bvolume{57}(\bissue{3}),
\bfpage{23}
(\byear{2024})
\doiurl{10.1007/s10686-024-09923-1}
{\href{https://arxiv.org/abs/2211.12052}{{arXiv:2211.12052}}}
{[astro-ph.HE]}
\end{barticle}
\endbibitem

\bibitem[\protect\citeauthoryear{Bhalerao et~al.}{2024b}]{bhalerao2024daksha}
\begin{barticle}
\bauthor{\bsnm{Bhalerao}, \binits{V.}},
\bauthor{\bsnm{Vadawale}, \binits{S.}},
\bauthor{\bsnm{Tendulkar}, \binits{S.}},
\bauthor{\bsnm{Bhattacharya}, \binits{D.}},
\bauthor{\bsnm{Rana}, \binits{V.}},
\bauthor{\bsnm{Adalja}, \binits{H.K.L.}},
\bauthor{\bsnm{Belatikar}, \binits{H.}},
\bauthor{\bsnm{Bhaganagare}, \binits{M.}},
\bauthor{\bsnm{Dewangan}, \binits{G.}},
\bauthor{\bsnm{Ghodgaonkar}, \binits{A.}}, \betal:
\batitle{{Daksha: on alert for high energy transients}}.
\bjtitle{Experimental Astronomy}
\bvolume{57}(\bissue{3}),
\bfpage{24}
(\byear{2024})
\doiurl{10.1007/s10686-024-09926-y}
{\href{https://arxiv.org/abs/2211.12055}{{arXiv:2211.12055}}}
{[astro-ph.IM]}
\end{barticle}
\endbibitem

\bibitem[\protect\citeauthoryear{{Catura} et~al.}{2000}]{catura2000performance}
\begin{bchapter}
\bauthor{\bsnm{{Catura}}, \binits{R.C.}},
\bauthor{\bsnm{{Bruner}}, \binits{M.E.}},
\bauthor{\bsnm{{Catura}}, \binits{P.R.}},
\bauthor{\bsnm{{Jurcevich}}, \binits{B.K.}},
\bauthor{\bsnm{{Kam}}, \binits{C.}},
\bauthor{\bsnm{{Lemen}}, \binits{J.R.}},
\bauthor{\bsnm{{Meyer}}, \binits{S.B.}},
\bauthor{\bsnm{{Morrison}}, \binits{M.D.}},
\bauthor{\bsnm{{Magida}}, \binits{M.B.}},
\bauthor{\bsnm{{Reid}}, \binits{P.B.}},
\bauthor{\bsnm{{Harvey}}, \binits{J.E.}},
\bauthor{\bsnm{{Thompson}}, \binits{P.L.}}:
\bctitle{{Performance of the engineering model x-ray mirror of the Solar X-ray
  Imager (SXI) for future GOES missions}}.
In: \beditor{\bsnm{{Hoover}}, \binits{R.B.}},
\beditor{\bsnm{{Walker}}, \binits{A.B.}} (eds.)
\bbtitle{X-Ray Optics, Instruments, and Missions IV}.
\bsertitle{Society of Photo-Optical Instrumentation Engineers (SPIE) Conference
  Series},
vol. \bseriesno{4138},
pp. \bfpage{33}--\blpage{42}
(\byear{2000}).
\doiurl{10.1117/12.407565}
\end{bchapter}
\endbibitem

\bibitem[\protect\citeauthoryear{{Conconi} et~al.}{2010}]{conconi2010wide}
\begin{barticle}
\bauthor{\bsnm{{Conconi}}, \binits{P.}},
\bauthor{\bsnm{{Campana}}, \binits{S.}},
\bauthor{\bsnm{{Tagliaferri}}, \binits{G.}},
\bauthor{\bsnm{{Pareschi}}, \binits{G.}},
\bauthor{\bsnm{{Citterio}}, \binits{O.}},
\bauthor{\bsnm{{Cotroneo}}, \binits{V.}},
\bauthor{\bsnm{{Proserpio}}, \binits{L.}},
\bauthor{\bsnm{{Civitani}}, \binits{M.}}:
\batitle{{A wide field X-ray telescope for astronomical survey purposes: from
  theory to practice}}.
\bjtitle{\mnras}
\bvolume{405}(\bissue{2}),
\bfpage{877}--\blpage{886}
(\byear{2010})
\doiurl{10.1111/j.1365-2966.2010.16513.x}
{\href{https://arxiv.org/abs/0912.5331}{{arXiv:0912.5331}}}
{[astro-ph.IM]}
\end{barticle}
\endbibitem

\bibitem[\protect\citeauthoryear{{Chen} et~al.}{2016}]{2016ChOpL..14l3401C}
\begin{barticle}
\bauthor{\bsnm{{Chen}}, \binits{S.}},
\bauthor{\bsnm{{Ma}}, \binits{S.}},
\bauthor{\bsnm{{Wang}}, \binits{Z.}}:
\batitle{{Wolter-I-like X ray telescope structure using one conical mirror and
  one quadric mirror}}.
\bjtitle{Chinese Optics Letters}
\bvolume{14}(\bissue{12}),
\bfpage{123401}--\blpage{123405}
(\byear{2016})
\doiurl{10.3788/COL201614.123401}
{\href{https://arxiv.org/abs/1608.02691}{{arXiv:1608.02691}}}
{[astro-ph.IM]}
\end{barticle}
\endbibitem

\bibitem[\protect\citeauthoryear{{Chase} and {van
  Speybroeck}}{1973}]{chase1973wolter}
\begin{barticle}
\bauthor{\bsnm{{Chase}}, \binits{R.C.}},
\bauthor{\bsnm{{van Speybroeck}}, \binits{L.P.}}:
\batitle{{Wolter-Schwarzschild telescopes for X-ray astronomy.}}
\bjtitle{\ao}
\bvolume{12},
\bfpage{1042}--\blpage{1044}
(\byear{1973})
\doiurl{10.1364/AO.12.001042}
\end{barticle}
\endbibitem

\bibitem[\protect\citeauthoryear{{De Korte} et~al.}{1981}]{de1981exosat}
\begin{barticle}
\bauthor{\bsnm{{De Korte}}, \binits{P.A.J.}},
\bauthor{\bsnm{{Giralt}}, \binits{R.}},
\bauthor{\bsnm{{Coste}}, \binits{J.N.}},
\bauthor{\bsnm{{Ernu}}, \binits{C.}},
\bauthor{\bsnm{{Frindel}}, \binits{S.}},
\bauthor{\bsnm{{Flamand}}, \binits{J.}},
\bauthor{\bsnm{{Contet}}, \binits{J.J.}}:
\batitle{{EXOSAT X-ray imaging optics.}}
\bjtitle{\ao}
\bvolume{20},
\bfpage{1080}--\blpage{1088}
(\byear{1981})
\doiurl{10.1364/AO.20.001080}
\end{barticle}
\endbibitem

\bibitem[\protect\citeauthoryear{{Elsner} et~al.}{2010}]{elsner2010methods}
\begin{bchapter}
\bauthor{\bsnm{{Elsner}}, \binits{R.F.}},
\bauthor{\bsnm{{O'Dell}}, \binits{S.L.}},
\bauthor{\bsnm{{Ramsey}}, \binits{B.D.}},
\bauthor{\bsnm{{Weisskopf}}, \binits{M.C.}}:
\bctitle{{Methods of optimizing x-ray optical prescriptions for wide-field
  applications}}.
In: \beditor{\bsnm{{Arnaud}}, \binits{M.}},
\beditor{\bsnm{{Murray}}, \binits{S.S.}},
\beditor{\bsnm{{Takahashi}}, \binits{T.}} (eds.)
\bbtitle{Space Telescopes and Instrumentation 2010: Ultraviolet to Gamma Ray}.
\bsertitle{Society of Photo-Optical Instrumentation Engineers (SPIE) Conference
  Series},
vol. \bseriesno{7732},
p. \bfpage{77322}
(\byear{2010}).
\doiurl{10.1117/12.856420}
\end{bchapter}
\endbibitem

\bibitem[\protect\citeauthoryear{{Giacconi}
  et~al.}{1979}]{giacconi1979einstein}
\begin{barticle}
\bauthor{\bsnm{{Giacconi}}, \binits{R.}},
\bauthor{\bsnm{{Branduardi}}, \binits{G.}},
\bauthor{\bsnm{{Briel}}, \binits{U.}},
\bauthor{\bsnm{{Epstein}}, \binits{A.}},
\bauthor{\bsnm{{Fabricant}}, \binits{D.}},
\bauthor{\bsnm{{Feigelson}}, \binits{E.}},
\bauthor{\bsnm{{Forman}}, \binits{W.}},
\bauthor{\bsnm{{Gorenstein}}, \binits{P.}},
\bauthor{\bsnm{{Grindlay}}, \binits{J.}},
\bauthor{\bsnm{{Gursky}}, \binits{H.}},
\bauthor{\bsnm{{Harnden}}, \binits{F.R.}},
\bauthor{\bsnm{{Henry}}, \binits{J.P.}},
\bauthor{\bsnm{{Jones}}, \binits{C.}},
\bauthor{\bsnm{{Kellogg}}, \binits{E.}},
\bauthor{\bsnm{{Koch}}, \binits{D.}},
\bauthor{\bsnm{{Murray}}, \binits{S.}},
\bauthor{\bsnm{{Schreier}}, \binits{E.}},
\bauthor{\bsnm{{Seward}}, \binits{F.}},
\bauthor{\bsnm{{Tananbaum}}, \binits{H.}},
\bauthor{\bsnm{{Topka}}, \binits{K.}},
\bauthor{\bsnm{{Van Speybroeck}}, \binits{L.}},
\bauthor{\bsnm{{Holt}}, \binits{S.S.}},
\bauthor{\bsnm{{Becker}}, \binits{R.H.}},
\bauthor{\bsnm{{Boldt}}, \binits{E.A.}},
\bauthor{\bsnm{{Serlemitsos}}, \binits{P.J.}},
\bauthor{\bsnm{{Clark}}, \binits{G.}},
\bauthor{\bsnm{{Canizares}}, \binits{C.}},
\bauthor{\bsnm{{Markert}}, \binits{T.}},
\bauthor{\bsnm{{Novick}}, \binits{R.}},
\bauthor{\bsnm{{Helfand}}, \binits{D.}},
\bauthor{\bsnm{{Long}}, \binits{K.}}:
\batitle{{The Einstein (HEAO 2) X-ray Observatory.}}
\bjtitle{\apj}
\bvolume{230},
\bfpage{540}--\blpage{550}
(\byear{1979})
\doiurl{10.1086/157110}
\end{barticle}
\endbibitem

\bibitem[\protect\citeauthoryear{{Garmire} et~al.}{2003}]{garmire2003advanced}
\begin{bchapter}
\bauthor{\bsnm{{Garmire}}, \binits{G.P.}},
\bauthor{\bsnm{{Bautz}}, \binits{M.W.}},
\bauthor{\bsnm{{Ford}}, \binits{P.G.}},
\bauthor{\bsnm{{Nousek}}, \binits{J.A.}},
\bauthor{\bsnm{{Ricker}}, \binits{J.} \bsuffix{George~R.}}:
\bctitle{{Advanced CCD imaging spectrometer (ACIS) instrument on the Chandra
  X-ray Observatory}}.
In: \beditor{\bsnm{{Truemper}}, \binits{J.E.}},
\beditor{\bsnm{{Tananbaum}}, \binits{H.D.}} (eds.)
\bbtitle{X-Ray and Gamma-Ray Telescopes and Instruments for Astronomy.}
\bsertitle{Society of Photo-Optical Instrumentation Engineers (SPIE) Conference
  Series},
vol. \bseriesno{4851},
pp. \bfpage{28}--\blpage{44}
(\byear{2003}).
\doiurl{10.1117/12.461599}
\end{bchapter}
\endbibitem

\bibitem[\protect\citeauthoryear{{Golub} et~al.}{2007}]{golub2007x}
\begin{barticle}
\bauthor{\bsnm{{Golub}}, \binits{L.}},
\bauthor{\bsnm{{DeLuca}}, \binits{E.}},
\bauthor{\bsnm{{Austin}}, \binits{G.}},
\bauthor{\bsnm{{Bookbinder}}, \binits{J.}},
\bauthor{\bsnm{{Caldwell}}, \binits{D.}},
\bauthor{\bsnm{{Cheimets}}, \binits{P.}},
\bauthor{\bsnm{{Cirtain}}, \binits{J.}},
\bauthor{\bsnm{{Cosmo}}, \binits{M.}},
\bauthor{\bsnm{{Reid}}, \binits{P.}},
\bauthor{\bsnm{{Sette}}, \binits{A.}}, \betal:
\batitle{{The X-Ray Telescope (XRT) for the Hinode Mission}}.
\bjtitle{\solphys}
\bvolume{243}(\bissue{1}),
\bfpage{63}--\blpage{86}
(\byear{2007})
\doiurl{10.1007/s11207-007-0182-1}
\end{barticle}
\endbibitem

\bibitem[\protect\citeauthoryear{{Giacconi} and
  {Rossi}}{1960}]{giacconi1960telescope}
\begin{barticle}
\bauthor{\bsnm{{Giacconi}}, \binits{R.}},
\bauthor{\bsnm{{Rossi}}, \binits{B.}}:
\batitle{{A `Telescope' for Soft X-Ray Astronomy}}.
\bjtitle{\jgr}
\bvolume{65},
\bfpage{773}
(\byear{1960})
\doiurl{10.1029/JZ065i002p00773}
\end{barticle}
\endbibitem

\bibitem[\protect\citeauthoryear{{Gaskin} et~al.}{2019}]{2019JATIS...5b1001G}
\begin{barticle}
\bauthor{\bsnm{{Gaskin}}, \binits{J.A.}},
\bauthor{\bsnm{{Swartz}}, \binits{D.A.}},
\bauthor{\bsnm{{Vikhlinin}}, \binits{A.}},
\bauthor{\bsnm{{{\"O}zel}}, \binits{F.}},
\bauthor{\bsnm{{Gelmis}}, \binits{K.E.}},
\bauthor{\bsnm{{Arenberg}}, \binits{J.W.}},
\bauthor{\bsnm{{Bandler}}, \binits{S.R.}},
\bauthor{\bsnm{{Bautz}}, \binits{M.W.}},
\bauthor{\bsnm{{Civitani}}, \binits{M.M.}},
\bauthor{\bsnm{{Dominguez}}, \binits{A.}},
\bauthor{\bsnm{{Eckart}}, \binits{M.E.}},
\bauthor{\bsnm{{Falcone}}, \binits{A.D.}},
\bauthor{\bsnm{{Figueroa-Feliciano}}, \binits{E.}},
\bauthor{\bsnm{{Freeman}}, \binits{M.D.}},
\bauthor{\bsnm{{G{\"u}nther}}, \binits{H.M.}},
\bauthor{\bsnm{{Havey}}, \binits{K.A.}},
\bauthor{\bsnm{{Heilmann}}, \binits{R.K.}},
\bauthor{\bsnm{{Kilaru}}, \binits{K.}},
\bauthor{\bsnm{{Kraft}}, \binits{R.P.}},
\bauthor{\bsnm{{McCarley}}, \binits{K.S.}},
\bauthor{\bsnm{{McEntaffer}}, \binits{R.L.}},
\bauthor{\bsnm{{Pareschi}}, \binits{G.}},
\bauthor{\bsnm{{Purcell}}, \binits{W.}},
\bauthor{\bsnm{{Reid}}, \binits{P.B.}},
\bauthor{\bsnm{{Schattenburg}}, \binits{M.L.}},
\bauthor{\bsnm{{Schwartz}}, \binits{D.A.}},
\bauthor{\bsnm{{Schwartz}}, \binits{E.D.}},
\bauthor{\bsnm{{Tananbaum}}, \binits{H.D.}},
\bauthor{\bsnm{{Tremblay}}, \binits{G.R.}},
\bauthor{\bsnm{{Zhang}}, \binits{W.W.}},
\bauthor{\bsnm{{Zuhone}}, \binits{J.A.}}:
\batitle{{Lynx X-Ray Observatory: an overview}}.
\bjtitle{Journal of Astronomical Telescopes, Instruments, and Systems}
\bvolume{5},
\bfpage{021001}
(\byear{2019})
\doiurl{10.1117/1.JATIS.5.2.021001}
\end{barticle}
\endbibitem

\bibitem[\protect\citeauthoryear{{Harvey} et~al.}{2005}]{2005SPIE.5867..114H}
\begin{bchapter}
\bauthor{\bsnm{{Harvey}}, \binits{J.E.}},
\bauthor{\bsnm{{Atanassova}}, \binits{M.}},
\bauthor{\bsnm{{Krywonos}}, \binits{A.}}:
\bctitle{{Systems engineering analysis of five 'as-manufactured' SXI
  telescopes}}.
In: \beditor{\bsnm{{Kahan}}, \binits{M.A.}} (ed.)
\bbtitle{Optical Modeling and Performance Predictions II}.
\bsertitle{Society of Photo-Optical Instrumentation Engineers (SPIE) Conference
  Series},
vol. \bseriesno{5867},
pp. \bfpage{114}--\blpage{124}
(\byear{2005}).
\doiurl{10.1117/12.622376}
\end{bchapter}
\endbibitem

\bibitem[\protect\citeauthoryear{{Harvey} et~al.}{2007}]{2007SPIE.6689E..0IH}
\begin{bchapter}
\bauthor{\bsnm{{Harvey}}, \binits{J.E.}},
\bauthor{\bsnm{{Krywonos}}, \binits{A.}},
\bauthor{\bsnm{{Atanassova}}, \binits{M.}},
\bauthor{\bsnm{{Thompson}}, \binits{P.L.}}:
\bctitle{{The Solar X-ray Imager on GOES-13: design, analysis, and on-orbit
  performance}}.
In: \beditor{\bsnm{{Fineschi}}, \binits{S.}},
\beditor{\bsnm{{Viereck}}, \binits{R.A.}} (eds.)
\bbtitle{Solar Physics and Space Weather Instrumentation II}.
\bsertitle{Society of Photo-Optical Instrumentation Engineers (SPIE) Conference
  Series},
vol. \bseriesno{6689},
p. \bfpage{66890}
(\byear{2007}).
\doiurl{10.1117/12.736870}
\end{bchapter}
\endbibitem

\bibitem[\protect\citeauthoryear{{Harvey} et~al.}{2001}]{harvey2001grazing}
\begin{barticle}
\bauthor{\bsnm{{Harvey}}, \binits{J.E.}},
\bauthor{\bsnm{{Krywonos}}, \binits{A.}},
\bauthor{\bsnm{{Thompson}}, \binits{P.L.}},
\bauthor{\bsnm{{Saha}}, \binits{T.T.}}:
\batitle{{Grazing-incidence hyperboloid hyperboloid designs for wide-field
  x-ray imaging applications}}.
\bjtitle{\ao}
\bvolume{40}(\bissue{1}),
\bfpage{136}--\blpage{144}
(\byear{2001})
\doiurl{10.1364/AO.40.000136}
\end{barticle}
\endbibitem

\bibitem[\protect\citeauthoryear{{Iizuka} et~al.}{2018}]{iizuka2018ground}
\begin{barticle}
\bauthor{\bsnm{{Iizuka}}, \binits{R.}},
\bauthor{\bsnm{{Hayashi}}, \binits{T.}},
\bauthor{\bsnm{{Maeda}}, \binits{Y.}},
\bauthor{\bsnm{{Ishida}}, \binits{M.}},
\bauthor{\bsnm{{Tomikawa}}, \binits{K.}},
\bauthor{\bsnm{{Sato}}, \binits{T.}},
\bauthor{\bsnm{{Kikuchi}}, \binits{N.}},
\bauthor{\bsnm{{Okajima}}, \binits{T.}},
\bauthor{\bsnm{{Soong}}, \binits{Y.}},
\bauthor{\bsnm{{Serlemitsos}}, \binits{P.J.}},
\bauthor{\bsnm{{Mori}}, \binits{H.}},
\bauthor{\bsnm{{Izumiya}}, \binits{T.}},
\bauthor{\bsnm{{Minami}}, \binits{S.}}:
\batitle{{Ground-based x-ray calibration of the Astro-H/Hitomi soft x-ray
  telescopes}}.
\bjtitle{Journal of Astronomical Telescopes, Instruments, and Systems}
\bvolume{4},
\bfpage{011213}
(\byear{2018})
\doiurl{10.1117/1.JATIS.4.1.011213}
\end{barticle}
\endbibitem

\bibitem[\protect\citeauthoryear{{Klimchuk}}{2006}]{2006SoPh..234...41K}
\begin{barticle}
\bauthor{\bsnm{{Klimchuk}}, \binits{J.A.}}:
\batitle{{On Solving the Coronal Heating Problem}}.
\bjtitle{\solphys}
\bvolume{234}(\bissue{1}),
\bfpage{41}--\blpage{77}
(\byear{2006})
\doiurl{10.1007/s11207-006-0055-z}
{\href{https://arxiv.org/abs/astro-ph/0511841}{{arXiv:astro-ph/0511841}}}
{[astro-ph]}
\end{barticle}
\endbibitem

\bibitem[\protect\citeauthoryear{{Lumb} et~al.}{2012}]{lumb2012x}
\begin{barticle}
\bauthor{\bsnm{{Lumb}}, \binits{D.H.}},
\bauthor{\bsnm{{Schartel}}, \binits{N.}},
\bauthor{\bsnm{{Jansen}}, \binits{F.A.}}:
\batitle{{X-ray Multi-mirror Mission (XMM-Newton) observatory}}.
\bjtitle{Optical Engineering}
\bvolume{51}(\bissue{1}),
\bfpage{011009}--\blpage{01100911}
(\byear{2012})
\doiurl{10.1117/1.OE.51.1.011009}
\end{barticle}
\endbibitem

\bibitem[\protect\citeauthoryear{{Lemen} et~al.}{2012}]{lemen2012atmospheric}
\begin{barticle}
\bauthor{\bsnm{{Lemen}}, \binits{J.R.}},
\bauthor{\bsnm{{Title}}, \binits{A.M.}},
\bauthor{\bsnm{{Akin}}, \binits{D.J.}},
\bauthor{\bsnm{{Boerner}}, \binits{P.F.}},
\bauthor{\bsnm{{Chou}}, \binits{C.}},
\bauthor{\bsnm{{Drake}}, \binits{J.F.}},
\bauthor{\bsnm{{Duncan}}, \binits{D.W.}},
\bauthor{\bsnm{{Edwards}}, \binits{C.G.}},
\bauthor{\bsnm{{Friedlaender}}, \binits{F.M.}},
\bauthor{\bsnm{{Heyman}}, \binits{G.F.}}, \betal:
\batitle{{The Atmospheric Imaging Assembly (AIA) on the Solar Dynamics
  Observatory (SDO)}}.
\bjtitle{\solphys}
\bvolume{275}(\bissue{1-2}),
\bfpage{17}--\blpage{40}
(\byear{2012})
\doiurl{10.1007/s11207-011-9776-8}
\end{barticle}
\endbibitem

\bibitem[\protect\citeauthoryear{{Marchesi} et~al.}{2020}]{2020A&A...642A.184M}
\begin{barticle}
\bauthor{\bsnm{{Marchesi}}, \binits{S.}},
\bauthor{\bsnm{{Gilli}}, \binits{R.}},
\bauthor{\bsnm{{Lanzuisi}}, \binits{G.}},
\bauthor{\bsnm{{Dauser}}, \binits{T.}},
\bauthor{\bsnm{{Ettori}}, \binits{S.}},
\bauthor{\bsnm{{Vito}}, \binits{F.}},
\bauthor{\bsnm{{Cappelluti}}, \binits{N.}},
\bauthor{\bsnm{{Comastri}}, \binits{A.}},
\bauthor{\bsnm{{Mushotzky}}, \binits{R.}},
\bauthor{\bsnm{{Ptak}}, \binits{A.}},
\bauthor{\bsnm{{Norman}}, \binits{C.}}:
\batitle{{Mock catalogs for the extragalactic X-ray sky: Simulating AGN surveys
  with ATHENA and with the AXIS probe}}.
\bjtitle{\aap}
\bvolume{642},
\bfpage{184}
(\byear{2020})
\doiurl{10.1051/0004-6361/202038622}
{\href{https://arxiv.org/abs/2008.09133}{{arXiv:2008.09133}}}
{[astro-ph.IM]}
\end{barticle}
\endbibitem

\bibitem[\protect\citeauthoryear{{Merloni} et~al.}{2012}]{merloni2012erosita}
\begin{botherref}
\oauthor{\bsnm{{Merloni}}, \binits{A.}},
\oauthor{\bsnm{{Predehl}}, \binits{P.}},
\oauthor{\bsnm{{Becker}}, \binits{W.}},
\oauthor{\bsnm{{B{\"o}hringer}}, \binits{H.}},
\oauthor{\bsnm{{Boller}}, \binits{T.}},
\oauthor{\bsnm{{Brunner}}, \binits{H.}},
\oauthor{\bsnm{{Brusa}}, \binits{M.}},
\oauthor{\bsnm{{Dennerl}}, \binits{K.}},
\oauthor{\bsnm{{Freyberg}}, \binits{M.}},
\oauthor{\bsnm{{Friedrich}}, \binits{P.}},
\oauthor{\bsnm{{Georgakakis}}, \binits{A.}},
\oauthor{\bsnm{{Haberl}}, \binits{F.}},
\oauthor{\bsnm{{Hasinger}}, \binits{G.}},
\oauthor{\bsnm{{Meidinger}}, \binits{N.}},
\oauthor{\bsnm{{Mohr}}, \binits{J.}},
\oauthor{\bsnm{{Nandra}}, \binits{K.}},
\oauthor{\bsnm{{Rau}}, \binits{A.}},
\oauthor{\bsnm{{Reiprich}}, \binits{T.H.}},
\oauthor{\bsnm{{Robrade}}, \binits{J.}},
\oauthor{\bsnm{{Salvato}}, \binits{M.}},
\oauthor{\bsnm{{Santangelo}}, \binits{A.}},
\oauthor{\bsnm{{Sasaki}}, \binits{M.}},
\oauthor{\bsnm{{Schwope}}, \binits{A.}},
\oauthor{\bsnm{{Wilms}}, \binits{J.}},
\oauthor{\bsnm{{German eROSITA Consortium}}, \binits{t.}}:
{eROSITA Science Book: Mapping the Structure of the Energetic Universe}.
arXiv e-prints,
1209--3114
(2012)
\doiurl{10.48550/arXiv.1209.3114}
{\href{https://arxiv.org/abs/1209.3114}{{arXiv:1209.3114}}}
{[astro-ph.HE]}
\end{botherref}
\endbibitem

\bibitem[\protect\citeauthoryear{{Mangus} and
  {Underwood}}{1969}]{mangus1969optical}
\begin{barticle}
\bauthor{\bsnm{{Mangus}}, \binits{J.D.}},
\bauthor{\bsnm{{Underwood}}, \binits{J.H.}}:
\batitle{{Optical design of a glancing incidence X-ray telescope.}}
\bjtitle{\ao}
\bvolume{8},
\bfpage{95}--\blpage{102}
(\byear{1969})
\doiurl{10.1364/AO.8.000095}
\end{barticle}
\endbibitem

\bibitem[\protect\citeauthoryear{{Mondal} et~al.}{2023}]{2023ApJ...955..146M}
\begin{barticle}
\bauthor{\bsnm{{Mondal}}, \binits{B.}},
\bauthor{\bsnm{{Vadawale}}, \binits{S.V.}},
\bauthor{\bsnm{{Del Zanna}}, \binits{G.}},
\bauthor{\bsnm{{Mithun}}, \binits{N.P.S.}},
\bauthor{\bsnm{{Sarkar}}, \binits{A.}},
\bauthor{\bsnm{{Mason}}, \binits{H.E.}},
\bauthor{\bsnm{{Janardhan}}, \binits{P.}},
\bauthor{\bsnm{{Bhardwaj}}, \binits{A.}}:
\batitle{{Evolution of Elemental Abundances in Hot Active Region Cores from
  Chandrayaan-2 XSM Observations}}.
\bjtitle{\apj}
\bvolume{955}(\bissue{2}),
\bfpage{146}
(\byear{2023})
\doiurl{10.3847/1538-4357/acdeeb}
{\href{https://arxiv.org/abs/2301.03519}{{arXiv:2301.03519}}}
{[astro-ph.SR]}
\end{barticle}
\endbibitem

\bibitem[\protect\citeauthoryear{{Mondal} et~al.}{2021}]{mondal2021darpanx}
\begin{barticle}
\bauthor{\bsnm{{Mondal}}, \binits{B.}},
\bauthor{\bsnm{{Vadawale}}, \binits{S.V.}},
\bauthor{\bsnm{{Mithun}}, \binits{N.P.S.}},
\bauthor{\bsnm{{Vaishnava}}, \binits{C.S.}},
\bauthor{\bsnm{{Tiwari}}, \binits{N.K.}},
\bauthor{\bsnm{{Goyal}}, \binits{S.K.}},
\bauthor{\bsnm{{Panini}}, \binits{S.S.}},
\bauthor{\bsnm{{Navalkar}}, \binits{V.}},
\bauthor{\bsnm{{Karmakar}}, \binits{C.}},
\bauthor{\bsnm{{Patel}}, \binits{M.R.}},
\bauthor{\bsnm{{Upadhyay}}, \binits{R.B.}}:
\batitle{{DarpanX: A python package for modeling X-ray reflectivity of
  multilayer mirrors}}.
\bjtitle{Astronomy and Computing}
\bvolume{34},
\bfpage{100446}
(\byear{2021})
\doiurl{10.1016/j.ascom.2020.100446}
{\href{https://arxiv.org/abs/2101.02571}{{arXiv:2101.02571}}}
{[astro-ph.IM]}
\end{barticle}
\endbibitem

\bibitem[\protect\citeauthoryear{{Nariai}}{1987}]{nariai1987geometrical}
\begin{barticle}
\bauthor{\bsnm{{Nariai}}, \binits{K.}}:
\batitle{{Geometrical aberration of a generalized Wolter type I telescope}}.
\bjtitle{\ao}
\bvolume{26}(\bissue{20}),
\bfpage{4428}--\blpage{4432}
(\byear{1987})
\doiurl{10.1364/AO.26.004428}
\end{barticle}
\endbibitem

\bibitem[\protect\citeauthoryear{{Nariai}}{1988}]{nariai1988geometric}
\begin{barticle}
\bauthor{\bsnm{{Nariai}}, \binits{K.}}:
\batitle{{Geometric aberration of a generalized Wolter type I telescope. 2:
  Analytical study}}.
\bjtitle{\ao}
\bvolume{27}(\bissue{2}),
\bfpage{345}--\blpage{350}
(\byear{1988})
\doiurl{10.1364/AO.27.000345}
\end{barticle}
\endbibitem

\bibitem[\protect\citeauthoryear{{Parker}}{1988}]{1988ApJ...330..474P}
\begin{barticle}
\bauthor{\bsnm{{Parker}}, \binits{E.N.}}:
\batitle{{Nanoflares and the Solar X-Ray Corona}}.
\bjtitle{\apj}
\bvolume{330},
\bfpage{474}
(\byear{1988})
\doiurl{10.1086/166485}
\end{barticle}
\endbibitem

\bibitem[\protect\citeauthoryear{Predehl et~al.}{2010}]{predehl2010erosita}
\begin{bchapter}
\bauthor{\bsnm{Predehl}, \binits{P.}},
\bauthor{\bsnm{B{\"o}hringer}, \binits{H.}},
\bauthor{\bsnm{Brunner}, \binits{H.}},
\bauthor{\bsnm{Brusa}, \binits{M.}},
\bauthor{\bsnm{Burwitz}, \binits{V.}},
\bauthor{\bsnm{Cappelluti}, \binits{N.}},
\bauthor{\bsnm{Churazov}, \binits{E.}},
\bauthor{\bsnm{Dennerl}, \binits{K.}},
\bauthor{\bsnm{Freyberg}, \binits{M.}},
\bauthor{\bsnm{Friedrich}, \binits{P.}}, \betal:
\bctitle{erosita on srg}.
In: \bbtitle{AIP Conference Proceedings},
vol. \bseriesno{1248},
pp. \bfpage{543}--\blpage{548}
(\byear{2010}).
\bcomment{American Institute of Physics}
\end{bchapter}
\endbibitem

\bibitem[\protect\citeauthoryear{{Pivovaroff} and
  {Okajima}}{2022}]{pivovaroff2023geometries}
\begin{bchapter}
\bauthor{\bsnm{{Pivovaroff}}, \binits{M.J.}},
\bauthor{\bsnm{{Okajima}}, \binits{T.}}:
\bctitle{{Geometries for Grazing Incidence Mirrors}}.
In: \beditor{\bsnm{{Bambi}}, \binits{C.}},
\beditor{\bsnm{{Sangangelo}}, \binits{A.}} (eds.)
\bbtitle{Handbook of X-ray and Gamma-ray Astrophysics},
p. \bfpage{115}
(\byear{2022}).
\doiurl{10.1007/978-981-16-4544-0_2-1}
\end{bchapter}
\endbibitem

\bibitem[\protect\citeauthoryear{{Saha}}{1986}]{saha1986transverse}
\begin{bchapter}
\bauthor{\bsnm{{Saha}}, \binits{T.T.}}:
\bctitle{{Transverse ray aberrations of Wolter type 1 telescopes.}}
In: \beditor{\bsnm{{Osantowski}}, \binits{J.F.}},
\beditor{\bsnm{{Van Speybroeck}}, \binits{L.P.}} (eds.)
\bbtitle{Grazing Incidence Optics}.
\bsertitle{Society of Photo-Optical Instrumentation Engineers (SPIE) Conference
  Series},
vol. \bseriesno{640},
pp. \bfpage{10}--\blpage{19}
(\byear{1986}).
\doiurl{10.1117/12.964352}
\end{bchapter}
\endbibitem

\bibitem[\protect\citeauthoryear{{Saha}}{1987}]{saha1987general}
\begin{barticle}
\bauthor{\bsnm{{Saha}}, \binits{T.T.}}:
\batitle{{General surface equations for glancing incidence telescopes}}.
\bjtitle{\ao}
\bvolume{26}(\bissue{4}),
\bfpage{658}--\blpage{663}
(\byear{1987})
\doiurl{10.1364/AO.26.000658}
\end{barticle}
\endbibitem

\bibitem[\protect\citeauthoryear{Serlemitsos et~al.}{1995}]{serlemitsos1995x}
\begin{barticle}
\bauthor{\bsnm{Serlemitsos}, \binits{P.J.}},
\bauthor{\bsnm{Jalota}, \binits{L.}},
\bauthor{\bsnm{Soong}, \binits{Y.}},
\bauthor{\bsnm{Kunieda}, \binits{H.}},
\bauthor{\bsnm{Tawara}, \binits{Y.}},
\bauthor{\bsnm{Tsusaka}, \binits{Y.}},
\bauthor{\bsnm{Suzuki}, \binits{H.}},
\bauthor{\bsnm{Sakima}, \binits{Y.}},
\bauthor{\bsnm{Yamazaki}, \binits{T.}},
\bauthor{\bsnm{Yoshioka}, \binits{H.}}, \betal:
\batitle{The x-ray telescope on board asca}.
\bjtitle{PASJ: Publications of the Astronomical Society of Japan (ISSN
  0004-6264), vol. 47, no. 1, p. 105-114}
\bvolume{47},
\bfpage{105}--\blpage{114}
(\byear{1995})
\end{barticle}
\endbibitem

\bibitem[\protect\citeauthoryear{{Shealy} and {Saha}}{1990}]{shealy1990formula}
\begin{barticle}
\bauthor{\bsnm{{Shealy}}, \binits{D.L.}},
\bauthor{\bsnm{{Saha}}, \binits{T.T.}}:
\batitle{{Formula for the rms blur circle radius of Wolter telescope based on
  aberration theory}}.
\bjtitle{\ao}
\bvolume{29}(\bissue{16}),
\bfpage{2433}--\blpage{2439}
(\byear{1990})
\doiurl{10.1364/AO.29.002433}
\end{barticle}
\endbibitem

\bibitem[\protect\citeauthoryear{{Schwartz} et~al.}{2019}]{schwartz2019lynx}
\begin{bchapter}
\bauthor{\bsnm{{Schwartz}}, \binits{D.A.}},
\bauthor{\bsnm{{Vikhlinin}}, \binits{A.}},
\bauthor{\bsnm{{Tananbaum}}, \binits{H.}},
\bauthor{\bsnm{{Freeman}}, \binits{M.}},
\bauthor{\bsnm{{Tremblay}}, \binits{G.}},
\bauthor{\bsnm{{Schwartz}}, \binits{E.D.}},
\bauthor{\bsnm{{Gaskin}}, \binits{J.A.}},
\bauthor{\bsnm{{Swartz}}, \binits{D.}},
\bauthor{\bsnm{{Gelmis}}, \binits{K.}},
\bauthor{\bsnm{{McCarley}}, \binits{K.S.}},
\bauthor{\bsnm{{Dominguez}}, \binits{A.}}:
\bctitle{{The Lynx X-ray Observatory: revealing the invisible universe}}.
In: \beditor{\bsnm{{Siegmund}}, \binits{O.H.}} (ed.)
\bbtitle{UV, X-Ray, and Gamma-Ray Space Instrumentation for Astronomy XXI}.
\bsertitle{Society of Photo-Optical Instrumentation Engineers (SPIE) Conference
  Series},
vol. \bseriesno{11118},
p. \bfpage{111180}
(\byear{2019}).
\doiurl{10.1117/12.2533637}
\end{bchapter}
\endbibitem

\bibitem[\protect\citeauthoryear{{Saha} and {Zhang}}{2022}]{saha2022optical}
\begin{barticle}
\bauthor{\bsnm{{Saha}}, \binits{T.T.}},
\bauthor{\bsnm{{Zhang}}, \binits{W.W.}}:
\batitle{{Optical design of type-1 x-ray telescopes and their application to
  STAR-X}}.
\bjtitle{\ao}
\bvolume{61}(\bissue{2}),
\bfpage{505}
(\byear{2022})
\doiurl{10.1364/AO.446958}
\end{barticle}
\endbibitem

\bibitem[\protect\citeauthoryear{{Saha} et~al.}{2014}]{2014SPIE.9144E..18S}
\begin{bchapter}
\bauthor{\bsnm{{Saha}}, \binits{T.T.}},
\bauthor{\bsnm{{Zhang}}, \binits{W.W.}},
\bauthor{\bsnm{{McClelland}}, \binits{R.S.}}:
\bctitle{{Optical design for a survey x-ray telescope}}.
In: \beditor{\bsnm{{Takahashi}}, \binits{T.}},
\beditor{\bsnm{{den Herder}}, \binits{J.-W.A.}},
\beditor{\bsnm{{Bautz}}, \binits{M.}} (eds.)
\bbtitle{Space Telescopes and Instrumentation 2014: Ultraviolet to Gamma Ray}.
\bsertitle{Society of Photo-Optical Instrumentation Engineers (SPIE) Conference
  Series},
vol. \bseriesno{9144},
p. \bfpage{914418}
(\byear{2014}).
\doiurl{10.1117/12.2055478}
\end{bchapter}
\endbibitem

\bibitem[\protect\citeauthoryear{Turner et~al.}{2001}]{turner2001european}
\begin{barticle}
\bauthor{\bsnm{Turner}, \binits{M.J.}},
\bauthor{\bsnm{Abbey}, \binits{A.}},
\bauthor{\bsnm{Arnaud}, \binits{M.}},
\bauthor{\bsnm{Balasini}, \binits{M.}},
\bauthor{\bsnm{Barbera}, \binits{M.}},
\bauthor{\bsnm{Belsole}, \binits{E.}},
\bauthor{\bsnm{Bennie}, \binits{P.}},
\bauthor{\bsnm{Bernard}, \binits{J.}},
\bauthor{\bsnm{Bignami}, \binits{G.}},
\bauthor{\bsnm{Boer}, \binits{M.}}, \betal:
\batitle{{The European Photon Imaging Camera on XMM-Newton: The MOS cameras}}.
\bjtitle{\aap}
\bvolume{365},
\bfpage{27}--\blpage{35}
(\byear{2001})
\doiurl{10.1051/0004-6361:20000087}
{\href{https://arxiv.org/abs/astro-ph/0011498}{{arXiv:astro-ph/0011498}}}
{[astro-ph]}
\end{barticle}
\endbibitem

\bibitem[\protect\citeauthoryear{{Thompson} and
  {Harvey}}{2000}]{thompson2000systems}
\begin{barticle}
\bauthor{\bsnm{{Thompson}}, \binits{P.L.}},
\bauthor{\bsnm{{Harvey}}, \binits{J.E.}}:
\batitle{{Systems engineering analysis of aplanatic Wolter type I x-ray
  telescopes}}.
\bjtitle{Optical Engineering}
\bvolume{39},
\bfpage{1677}--\blpage{1691}
(\byear{2000})
\doiurl{10.1117/1.602545}
\end{barticle}
\endbibitem

\bibitem[\protect\citeauthoryear{{Tiwari} et~al.}{2024}]{tiwari2024darsakx}
\begin{barticle}
\bauthor{\bsnm{{Tiwari}}, \binits{N.K.}},
\bauthor{\bsnm{{Vadawale}}, \binits{S.V.}},
\bauthor{\bsnm{{Mithun}}, \binits{N.P.S.}},
\bauthor{\bsnm{{Vaishnava}}, \binits{C.S.}},
\bauthor{\bsnm{{Saiguhan}}, \binits{B.}}:
\batitle{{DarsakX: A Python package for designing and analyzing imaging
  performance of X-ray telescopes}}.
\bjtitle{Astronomy and Computing}
\bvolume{47},
\bfpage{100829}
(\byear{2024})
\doiurl{10.1016/j.ascom.2024.100829}
{\href{https://arxiv.org/abs/2405.06343}{{arXiv:2405.06343}}}
{[astro-ph.IM]}
\end{barticle}
\endbibitem

\bibitem[\protect\citeauthoryear{{van Speybroeck} and
  {Chase}}{1972}]{vanspeybroeck1972design}
\begin{barticle}
\bauthor{\bsnm{{van Speybroeck}}, \binits{L.P.}},
\bauthor{\bsnm{{Chase}}, \binits{R.C.}}:
\batitle{{Design parameters of paraboloid-hyperboloid telescopes for X-ray
  astronomy.}}
\bjtitle{\ao}
\bvolume{11},
\bfpage{440}--\blpage{445}
(\byear{1972})
\doiurl{10.1364/AO.11.000440}
\end{barticle}
\endbibitem

\bibitem[\protect\citeauthoryear{{Weisskopf}}{2012}]{weisskopf2012chandra}
\begin{barticle}
\bauthor{\bsnm{{Weisskopf}}, \binits{M.C.}}:
\batitle{{Chandra x-ray optics}}.
\bjtitle{Optical Engineering}
\bvolume{51}(\bissue{1}),
\bfpage{011013}--\blpage{0110138}
(\byear{2012})
\doiurl{10.1117/1.OE.51.1.011013}
{\href{https://arxiv.org/abs/1110.4020}{{arXiv:1110.4020}}}
{[astro-ph.IM]}
\end{barticle}
\endbibitem

\bibitem[\protect\citeauthoryear{{Werner}}{1977}]{werner1977imaging}
\begin{barticle}
\bauthor{\bsnm{{Werner}}, \binits{W.}}:
\batitle{{Imaging properties of Wolter I type X-ray telescopes.}}
\bjtitle{\ao}
\bvolume{16},
\bfpage{764}--\blpage{773}
(\byear{1977})
\doiurl{10.1364/AO.16.000764}
\end{barticle}
\endbibitem

\bibitem[\protect\citeauthoryear{{Windt}}{2015}]{windt2015advancements}
\begin{bchapter}
\bauthor{\bsnm{{Windt}}, \binits{D.L.}}:
\bctitle{{Advancements in hard x-ray multilayers for x-ray astronomy}}.
\bsertitle{Society of Photo-Optical Instrumentation Engineers (SPIE) Conference
  Series},
vol. \bseriesno{9603},
p. \bfpage{96031}
(\byear{2015}).
\doiurl{10.1117/12.2187481}
\end{bchapter}
\endbibitem

\bibitem[\protect\citeauthoryear{Wolter}{1952a}]{wolter1952spiegelsysteme}
\begin{barticle}
\bauthor{\bsnm{Wolter}, \binits{H.}}:
\batitle{Spiegelsysteme streifenden einfalls als abbildende optiken f{\"u}r
  r{\"o}ntgenstrahlen}.
\bjtitle{Annalen der Physik}
\bvolume{445}(\bissue{1-2}),
\bfpage{94}--\blpage{114}
(\byear{1952})
\end{barticle}
\endbibitem

\bibitem[\protect\citeauthoryear{Wolter}{1952b}]{wolter1952verallgemeinerte}
\begin{barticle}
\bauthor{\bsnm{Wolter}, \binits{H.}}:
\batitle{Verallgemeinerte schwarzschildsche spiegelsysteme streifender
  reflexion als optiken f{\"u}r r{\"o}ntgenstrahlen}.
\bjtitle{Annalen der Physik}
\bvolume{445}(\bissue{4-5}),
\bfpage{286}--\blpage{295}
(\byear{1952})
\end{barticle}
\endbibitem

\bibitem[\protect\citeauthoryear{Wolter}{1971}]{wolter1971bildfehlerabschatzung}
\begin{barticle}
\bauthor{\bsnm{Wolter}, \binits{H.}}:
\batitle{Bildfehlerabsch{\"a}tzung f{\"u}r r{\"o}ntgenstrahlenteleskope}.
\bjtitle{Optica Acta: International Journal of Optics}
\bvolume{18}(\bissue{6}),
\bfpage{425}--\blpage{429}
(\byear{1971})
\end{barticle}
\endbibitem

\bibitem[\protect\citeauthoryear{{Zhang}}{2009}]{zhang2009manufacture}
\begin{bchapter}
\bauthor{\bsnm{{Zhang}}, \binits{W.W.}}:
\bctitle{{Manufacture of mirror glass substrates for the NuSTAR mission}}.
In: \beditor{\bsnm{{O'Dell}}, \binits{S.L.}},
\beditor{\bsnm{{Pareschi}}, \binits{G.}} (eds.)
\bbtitle{Optics for EUV, X-Ray, and Gamma-Ray Astronomy IV}.
\bsertitle{Society of Photo-Optical Instrumentation Engineers (SPIE) Conference
  Series},
vol. \bseriesno{7437},
p. \bfpage{74370}
(\byear{2009}).
\doiurl{10.1117/12.830225}
\end{bchapter}
\endbibitem

\end{thebibliography}

\end{document}